\documentclass[aps,pra,twocolumn]{revtex4-2}
\usepackage{latexsym,amssymb,amsmath,graphicx,mathtools}
\usepackage[dvipsnames]{xcolor}
\usepackage[noabbrev]{cleveref}
\usepackage{physics}
\usepackage{amsthm}
    \newtheorem{result}{Result}

\usepackage[caption=false]{subfig}

\DeclareGraphicsExtensions{.png, .pdf,.jpg}

\newcommand{\state}{\hat{\rho}}

\newcommand{\q}{\hat{X}^\pm}
\newcommand{\qx}{\hat{X}^+}
\newcommand{\qp}{\hat{X}^-}

\renewcommand{\vec}[1]{\mathbf{#1}}
\newcommand{\mat}[1]{\mathbf{#1}}
\newcommand{\mean}{\vec{\bar{r}}}
\newcommand{\cov}{\mat{V}}

\newcommand{\vvariance}[2]{\left\langle\left\{ #1 \, ; \, #2 \right\} \right\rangle }
\newcommand{\expectation}[1]{\langle#1\rangle}

\newcommand{\reals}{\mathbb{R}}

\newcommand{\e}{\mathrm{e}}

\newcommand{\steering}{E_{1|2}(g)}

\begin{document}

\title{Quantum steering as a resource for secure tripartite quantum state sharing}
\author{Cailean Wilkinson, Matthew Thornton and Natalia Korolkova}
\affiliation{School of Physics and Astronomy, University of St Andrews,
North Haugh, St Andrews KY16 9SS, UK}
\begin{abstract}
	Quantum state sharing (QSS) is a protocol by which a (secret) quantum state may be securely split into shares, shared between multiple potentially-dishonest players, and reconstructed. Crucially the players are each assumed to be dishonest, and so QSS requires that only a collaborating authorised subset of players can access the original secret state; any dishonest unauthorised conspiracy cannot reconstruct it. We analyse a QSS protocol involving three untrusted players and demonstrate that \emph{quantum steering} is the required resource enabling the protocol to proceed securely. We analyse the level of steering required to share any single-mode Gaussian secret which enables the states to be shared with the optimal use of resources.
\end{abstract}
\maketitle

\section{Introduction} \label{sec:introduction}
Secret sharing is a cryptographic process which splits information between several players such that it is inaccessible to any individual player but can be accessed when players collaborate \cite{shamir_1979,blakley_1979,schneier_1996}. By requiring collaboration, secret sharing provides guaranteed security against small groups of dishonest actors. Secret-sharing schemes might be used, for example, by a bank manager to share the vault combination between their staff such that a number of them are required to access it.

Quantum state sharing (QSS) \footnote{As distinct from quantum secret sharing, which uses quantum resources to securely share classical information \cite{richter_2021}.} translates this scheme to act on quantum secrets: the information describing a single quantum state (not known to the dealer) is shared between the modes of a larger multipartite system. Since no individual mode contains enough information to reconstruct the original state, only certain authorised subsets of players can access the original state through collaboration \cite{cleve_1999, hillery_1999}. As in the classical case, quantum state sharing then provides security against small groups of dishonest parties.

This class of protocol aims at a variety of future uses in diverse quantum technology schemes. In secure distributed quantum computing, computations could be performed on each share individually without any single quantum computer having direct access to the underlying information \cite{ouyang_2017}. The shares from different computers could then be recombined to produce the computation outcome. This form of so-called blind quantum computation allows untrusted quantum computers to be used securely without fear of data loss. Quantum state sharing may also find use in loss-tolerant quantum information distribution as a form of quantum error correction analogous to Reed-Solomon codes \cite{grassl_1999}, potentially forming a crucial building block to a future quantum internet \cite{wehner_2018} or finding uses within a quantum computing stack \cite{ball_2021}. 

The prototypical quantum-state-sharing scheme is $(k,n)$-threshold QSS, in which the secret state is split into $n$ shares with a predefined threshold number of shares $k$ required to reconstruct it. Any subset of shares meeting this threshold can then reconstruct the state. While this may initially present as a limited form of QSS, more complex schemes - in which a different number of shares is required depending on which shares are involved - can be built simply by distributing an uneven number of shares to each player. The primary restriction on this approach is common to any QSS scheme: to avoid breaking the no-cloning theorem a reconstruction is not possible with fewer than half of all shares. For illustration, in this paper we will consider the simplest non-trivial case: {$(2,3)$-threshold} QSS in which any two of a total of three shares may reconstruct the original state.

Quantum state sharing was first formalised in the continuous-variable regime by Tyc and Sanders \cite{tyc_2002} with some possible implementations of $(2,3)$-threshold QSS utilising two-mode squeezed-state resources later demonstrated and discussed by Lance \textit{et al.} \cite{lance_2003,lance_2004,lance_2005}. In this paper, we present a generalised version of Lance \textit{et al.}'s protocol which allows for the use of \textit{any generally asymmetric Gaussian resource state}. The previous protocol can be obtained as a special symmetric case of the one presented here. In contrast to previous works, we model the reconstruction process simply as a quantum channel, leaving the choice of physical implementation free. Finally, we also consider the use of this scheme for the sharing of \textit{any} arbitrary single-mode Gaussian state, providing a complete image of tripartite Gaussian QSS.

For any quantum information task to be useful in a real-world setting, it must provide guaranteed security. For quantum state sharing, this means that the honest collaborating parties must be able to reconstruct a better copy of the original state than any adversaries in every case. With perfect entanglement, this protocol is secure for the sharing of any single-mode Gaussian state. However, increasing entanglement requires greater quantum resources and adds cost to the implementation. With the quantum technologies era emerging, the thrifty and careful use of these quantum resources is becoming imperative. To that end, we analyse here the minimum required conditions under which a fully Gaussian $(2,3)$-threshold QSS can be considered secure and demonstrate that quantum steering is the resource required. In particular, we show that any two-mode state which is one way steerable can be used as a resource to securely share a coherent state, and we analyse the strength of steering required to share a general single-mode Gaussian state. In previous discussion of the security of continuous-variable QSS \cite{lance_2003}, the security was derived from the inclusion of classically-correlated Gaussian noise in the shares. While such noise can reduce the amount of information obtainable from a single share to an arbitrarily small degree, it leaves the protocol vulnerable to eavesdropping in the classical noise generation stage in a way that a fully quantum approach does not. In this paper, we consider only security derived from the no-cloning theorem, which can be further augmented by the inclusion of classical noise but is not reliant on it.

After briefly reviewing the entanglement properties of two-mode Gaussian states in \cref{sec:resources}, we outline the details of the $(2,3)$-threshold quantum-state-sharing scheme under discussion in \cref{sec:qss_overview,sec:unitygain}. We then discuss the security of the protocol for coherent states in \cref{sec:security} and for general single-mode Gaussian states in \cref{sec:other_states}. Additional technical results may be found in the Supplemental Material.

\section{Gaussian resource states} \label{sec:resources}
We begin with a brief review of the properties of entangled two-mode Gaussian states, which form the resource for this protocol. At this stage, we first wish to clarify the sense in which we use the term ``resource'' in this paper. We depart from the formal definition used in resource theory \cite{chitambar_2019} of a property that cannot be created at will by the participants using only local operations. Under that definition, this protocol would require no resource as the resource-state preparation could equivalently be absorbed into the protocol. Instead, we take a looser, more experiment-inspired definition of the resource as that property which enables the quantum advantage - and thus the resource state as that state which provides this property.

 A Gaussian state is one whose Wigner function is Gaussian and, consequently, is fully characterised by its mean vector $\mean\in\reals^{2n}$ and covariance matrix $\cov\in\reals^{2n\times2n}$ \cite{weedbrook_2012}. We define elements of the covariance matrix as $\cov_{i,j} = \vvariance{\Delta^2\hat X_i}{\Delta^2\hat X_j}$ where $\{...\}$ represents the anticommutator, $\Delta^2(\hat{O})=\expectation{\hat{O}^2} - \expectation{\hat{O}}^2$ represents the variance of operator $\hat{O}$, and $\qx = (\hat{a}^\dagger+\hat{a})/\sqrt{2}$ and $\qp = i(\hat{a}^\dagger-\hat{a})/\sqrt{2}$ represent the position $\qx$ and the momentum $\qp$ quadratures of each mode, respectively.

As we show in \cref{sec:security}, secure QSS requires a strict form of entanglement in which the measurement of one mode can affect the state of the second mode. This is known as Einstein-Podolsky-Rosen (EPR) steering \cite{wiseman_2007,reid_1989}. The ability of one mode of a two-mode state to EPR steer the other is quantified through the steering parameter \cite{he_2015a}
\begin{align}
    E_{1|2}(\vec{g}) = \Delta(\qx_1 - g^+\qx_2)\Delta(\qp_1 + g^-\qp_2),
\end{align}
where $\q_i$ represents the quadrature operators for each mode and $\Delta(\hat{O}) \coloneqq \sqrt{\Delta^2(\hat{O})} = \sqrt{\expectation{\hat{O}^2} - \expectation{\hat{O}}^2}$ represents the square root of the variance of $\hat{O}$. Mode $2$ can steer mode $1$ whenever there exists a $\vec{g}=(g^+,g^-)^\mathrm{T}\in\reals^2$ such that $E_{1|2}(\vec{g})<1$ with greater EPR steering as $E_{1|2}\rightarrow0$.  Notably, this quantity is directional so a state may be steerable from mode $2\rightarrow1$ but not from $1\rightarrow2$. A state is said to be two-way steerable when it is steerable in both directions.

The steering parameter measures the correlation between $\q_1$ and $g^\pm\q_2$, in which $g^\pm$ represents an effective scaling between the modes. Equivalently, $E_{1|2}(\vec{g})$ represents the extent to which the two resource modes cancel when mixed as
\begin{align}
    \qx_1 - g^+ \qx_2, 
    \qquad\qquad
    \qp_1 + g^- \qp_2,
\end{align}
quantified by the smallness of the corresponding uncertainties.

For simplicity, we restrict our discussion to those resource states which exhibit equivalent entanglement properties in each quadrature, the $(X-P)$-balanced states. For such states, the steering parameter reduces to
\begin{align}
    E_{1|2}(g) = \Delta^2(\qx_1 - g\qx_2) = \Delta^2(\qp_1 + g\qp_2),
    \label{eqn:steering}
\end{align}
for $g^+=g^-:=g$. Although this condition is presented here for any $g\in\reals$, it can be shown that steering is only possible for $g\in(0,\sqrt{2})$, and so it is this range that we will consider in this paper \footnote{This can be seen by considering the channel resulting in $\q_\text{out}=1/\sqrt{1-g^2}(\q_{r1} \mp g\q_{r2})$ for $g<1$ and imposing the uncertainty limit on this output to get a condition for minimum $\steering>1-g^2$. This can be converted to a condition on steering in the opposite direction for $\bar{g}=1/g>1$ of $E_{2|1}(\bar{g})>\bar{g}^2-1$ and so steering can only be certified ($\steering<1$) for values of $g\in(0,\sqrt{2})$.}. One common example of an $(X-P)$-balanced state is the two-mode squeezed vacuum state with squeezing $\zeta$, which has steering parameter
\begin{align}
    E_{1|2}(g) = (1+g^2) \cosh(2\zeta) - 2g\sinh(2\zeta).
\end{align}

\section{Tripartite quantum state sharing} \label{sec:qss_overview}
\begin{figure}[ht] 
\centering
\includegraphics[width=\linewidth]{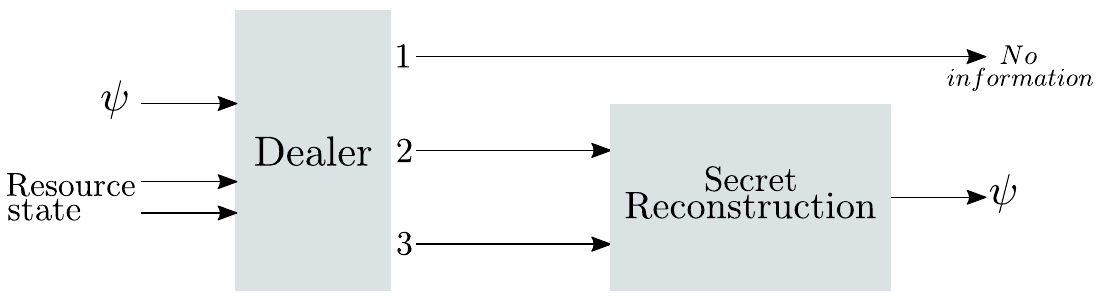}
\caption{Quantum-state-sharing schemes consist of two distinct sub-protocols. In $\left(2,3\right)$-threshold QSS, the secret state $\psi$ is originally passed to a dealer who mixes it with an entangled resource state to produce three shares (dealer protocol), none of which contain a suitable amount of information about the secret state. Any two of these shares can then be recombined with a suitable reconstruction protocol to recover the original secret state.}
\label{fig:setup}
\end{figure}
\begin{figure*}
    \subfloat[][]{
        \includegraphics[width=0.3\textwidth]{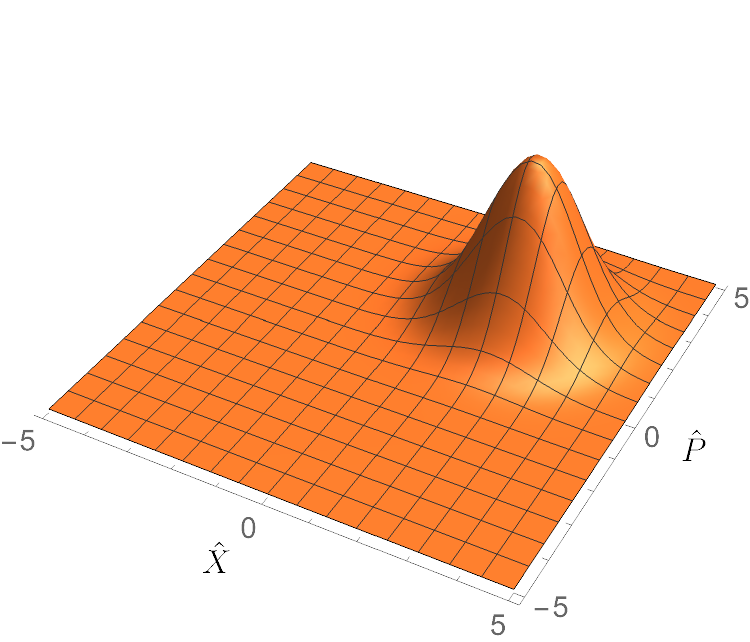}
    }
    \subfloat[][]{
        \includegraphics[width=0.3\textwidth]{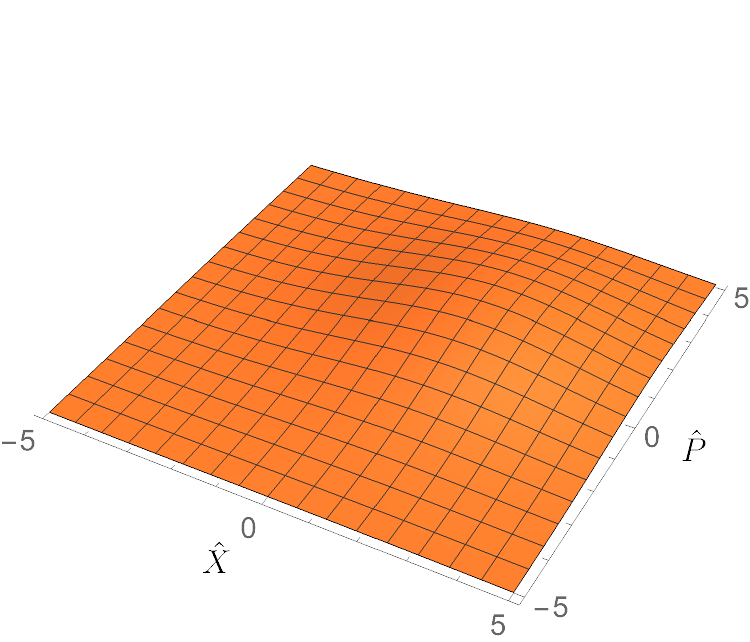}
    }
    \subfloat[][]{
        \includegraphics[width=0.3\textwidth]{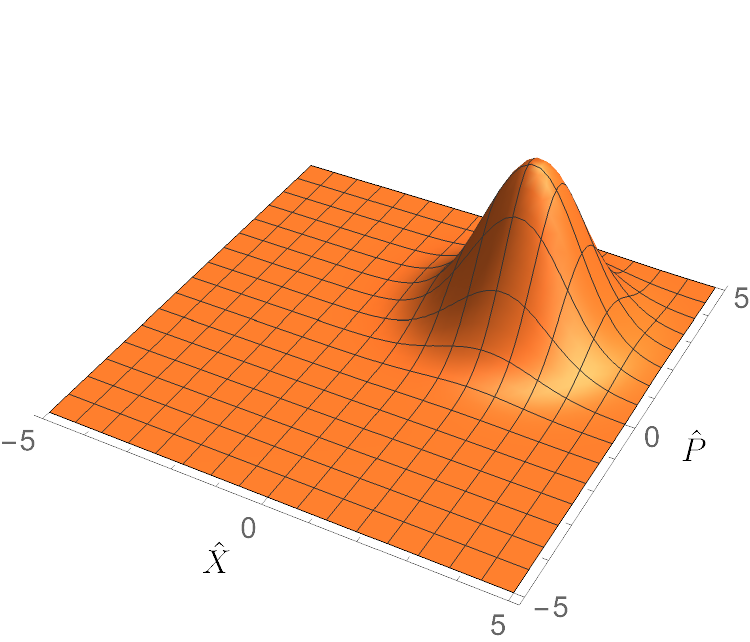}
    }
    \caption{Wigner functions at each stage of the QSS process: (a) input state, (b) intermediate distributed share, and (c) reconstructed output state. A single intermediate share does not contain enough information for a dishonest party to recover the original state. By combining any two shares, however, the original state can be reconstructed to a reasonable fidelity, as seen by comparing (c) to (a). This figure shows the sharing of a coherent-state secret with mean $\mean=(2,2)^\mathrm{T}$ utilising a two-mode squeezed vacuum resource state with 13dB squeezing.}

    \label{fig:wigner}
\end{figure*}

We now turn our attention to the specific QSS protocol we are interested in. Threshold quantum state sharing consists of two distinct stages: the \emph{dealer protocol}, in which the single-mode secret is split into multiple shares; and the \emph{reconstruction protocol}, in which a subset of these shares are recombined to reproduce the original secret state. Neither of these participants has any knowledge about the secret state. In $(2,3)$-threshold QSS, the dealer mixes the secret state with one mode of the two-mode resource state to produce an entangled system of three modes. Any two of these modes can then be used to reconstruct the original secret state, through a sub-protocol which we denote \textit{$\{i,j\}$ reconstruction} when modes $i$ and $j$ are used. An overview of the QSS protocol is shown in \cref{fig:setup}, and an illustration of the Wigner functions representing each stage of the process is shown in \cref{fig:wigner}.


\paragraph{Dealer protocol.} The dealer constructs the three shares by interfering the secret state on a balanced beam splitter with one mode of the two-mode entangled resource. The three output states are then related to the input states by
\begin{align}
    \q_1 &= \frac{1}{\sqrt{2}} ( \q_\psi + \q_{r1} ), \\
    \q_2 &= \frac{1}{\sqrt{2}} ( \q_\psi - \q_{r1} ), \\
    \q_3 &= \q_{r2},
\end{align}
where $\q_\psi$ are the quadrature operators representing the secret state and $\q_{ri}$ represents each mode of the resource state. Crucially, none of $\q_{1,2,3}$ individually contain enough information to accurately reconstruct the original state. As shown in Ref. \cite{lance_2004}, the information obtainable from each share could be further reduced with the addition of correlated classical noise without impacting the reconstructed state. Each of the three modes are distributed to a player as labelled.

\paragraph{\textit{\{1,2\} reconstruction}.}\label{sec:23reconstruction} Players $1$ and $2$ may trivially reconstruct their state by passing each share through a second balanced beam splitter. This will effectively reverse the beam splitter used in the dealer protocol, reproducing the original separable system and leaving in one of the beam-splitter outputs the state
\begin{align}
    \q_\text{out} &= \frac{1}{\sqrt{2}} \left( \q_1 + \q_2 \right) = \q_\psi.
\end{align}
In the ideal case, with no transmission or component losses, $\q_\psi$ will be reconstructed perfectly regardless of the resource state used. 

\paragraph{\textit{\{1,3\}} and \textit{\{2,3\} reconstruction}:}
Reconstructing the original state using share 3 requires a more complex disentanglement process to separate $\q_\psi$ from the resource state. We focus here on $\{1,3\}$ reconstruction; the $\{2, 3\}$ case follows with only minor changes. Recalling from \cref{eqn:steering} that the resource state is entangled such that the modes cancel maximally when mixed with ratio $\q_{r1} \mp g\q_{r2}$ for some $g\in\reals$, it becomes clear that in order to recover $\q_\psi$ we wish to implement the transformation
\begin{align} \label{eqn:reconstruction_13}
    \q_\text{out} &\rightarrow \eta \left[ \q_\psi + (\q_{r1} \mp g\q_{r2}) \right] \\
    &= \eta \left[\sqrt{2}\q_1 \mp g\q_3 \right]\notag,
\end{align}
where $\eta\in\reals$ represents an amplification of the output state which is required to preserve the canonical commutation relations. The players can control $g$ by adjusting the parameters of the reconstruction process (see the supplemental material for an example setup \cite{appendix}), and so its value also acts as a unique label for a specific reconstruction setup.

This transformation produces a state with mean $\mean = \eta\mean_\psi$ and covariance matrix $\cov = \eta^2\cov_\psi + \eta^2\steering \mat{I}$: an amplified, generally-noisy copy of the input state. We show in the Supplemental Material \cite{appendix} that to preserve the canonical commutation relations, and thus satisfy the uncertainty theorem, this reconstruction must impose a gain of $\eta = 1/\sqrt{2-g^2}$ on the secret state. Clearly, then, this reconstruction protocol amplifies the original state for all $g>1$ and de-amplifies it for all $g<1$ - the original state is reproduced with unity gain only for $g=1$.

\section{Unity-gain quantum state sharing} \label{sec:unitygain}
When the protocol is implemented for $g=1$, the output state $\state_{\text{out}}$ is reproduced with the same mean $\mean_\psi$ as the secret state and with covariance matrix $\cov_\text{out} = \cov_\psi + E_{1|2}\left(g=1\right)\mat{I}$. The accuracy of this reconstruction can be quantified by the fidelity $\mathcal{F}=\bra{\psi}\state_\text{out}\ket{\psi}$ between the original secret state $\psi$ and the output state. When the output and input states have the same mean amplitude $\mean$, this fidelity can be expressed in terms of the covariance matrices as $\mathcal{F}=2/\sqrt{\det(\cov_\psi+\cov_\text{out})}$. The ideal fidelity for QSS implemented for $g=1$ is then
\begin{align}
    \mathcal{F}_{g=1} = \frac{2}{2+E_{1|2}\left(g=1\right)}.
\end{align}

In general, for $g\ne1$, the protocol as outlined in \cref{sec:qss_overview} will not reproduce the mean $\mean$ of the input state exactly. To correct for the change in $\mean$ introduced by the protocol, and so reconstruct the original state with unity gain, we augment it with an additional pre-amplification or post-attenuation step. These are corrections similar to those introduced for quantum teleportation in Ref.~\cite{he_2015} and so we also describe them as \textit{late-stage attenuation (lsatt)} and \textit{early-stage amplification (esa)} QSS.

\subsection{Late-stage attenuation \textit{(lsatt)}} When the output state is an amplified copy of the input state (when $\eta>1$, $g>1$), the optimum correction is to attenuate the output state after the QSS reconstruction protocol. Modelling this attenuation as an ideal beam splitter with transmissivity $\tau = 1/\eta^2$ with a vacuum environment implements the transformation $\q_\text{out} \rightarrow \frac{1}{\eta} \q_\text{out} + \sqrt{1-\frac{1}{\eta^2}}\q_\text{vac}$. The corrected output state then has mean $\mean_\text{out} = \mean_\text{in}$ and covariance matrix $\cov_\text{out} = \cov_\text{in} + (E_{1|2}(g) + 1 - 1/\eta^2)\mat{I}$. 

For the specific case of a coherent-state secret with covariance matrix $\cov_\text{in} = \mat{I}$, the secret state is reproduced with a fidelity of
\begin{align}\label{eqn:lsatt_fidelity}
    \mathcal{F}_\textit{lsatt} = \frac{2}{3 - 1/\eta^2 + E_{1|2}\left(g\right)}.
\end{align}

\subsection{Early-stage amplification \textit{(esa)}}
When the output state is a de-amplified copy of the input state (when $\eta<1$, $g<1$), the optimum correction is instead to amplify the input secret state prior to the dealer protocol. We model this process as an ideal amplifying channel; in practice such an amplification could be achieved by a phase-insensitive amplifier \cite{weedbrook_2012}. Denoting the original secret state by $\psi$, the amplified input to the QSS protocol can be written $\q_\text{in} = \frac{1}{\eta}\q_\psi + \sqrt{\frac{1}{\eta^2} - 1}\q_\text{vac}$ where $1/\eta>1$. Following the de-amplifying QSS protocol, the output state will have mean $\mean_\text{out}=\eta\mean_\text{in}=\mean_\psi$ and covariance matrix $\cov_\text{out} = \eta^2 \cov_\text{in} + \eta^2 E_{1|2}(g)\mat{I} = \cov_\psi + (\eta^2 E_{1|2}\left(g\right) + 1 - \eta^2)\mat{I}$. 

For a coherent-state secret, the secret state is reproduced with a fidelity of
\begin{align}\label{eqn:esa_fidelity}
    \mathcal{F}_\textit{esa} = \frac{2}{3 - \eta^2 + \eta^2 E_{1|2}\left(g\right)}
\end{align}

Of course, the introduction of an amplification stage prior to the dealer protocol would require a corresponding de-amplification correction for $\left\{1, 2\right\}$-reconstruction. However, since, under equal conditions, $\left\{1, 2\right\}$-reconstruction will always have higher fidelity than $\left\{2, 3\right\}$ or $\left\{1,3\right\}$-reconstruction, this would not affect our analysis of the security of the protocol.

\section{Security analysis for coherent-state QSS} \label{sec:security}
For a quantum-state-sharing scheme to be considered secure it must be guaranteed that the collaborating parties obtain more information about the original secret than any adversary can. This security requirement can be certified through the uncertainty theorem, which imposes that only one copy of a single quantum state can exceed a fidelity of $\mathcal{F} = 2/3$ - a condition termed the no-cloning limit \cite{grosshans_2001}. Should the collaborators reconstruct the state with fidelity above this limit, it follows immediately that no other party can obtain as much information as them and so the protocol is secure. The optimal fidelity $\mathcal{F} = 2/3$ for duplication of a coherent state ($1 \rightarrow 2$ cloning) and its extension to $N \rightarrow M$ cloning has been studied in \cite{cerf_2000}. In this paper we assume any eavesdroppers to be limited to Gaussian operations; it has been shown that with the use of non-Gaussian operations, coherent states can be cloned with fidelity up to $\mathcal{F}\approx0.68$, and so loosening this assumption would slightly increase the following entanglement requirements \cite{cerf_2005}.

 To certify security for the whole protocol, each possible reconstruction (\textit{\{1,2\}}, \textit{\{1,3\}}, \textit{\{2,3\}}) must individually be provably secure. Since the reconstruction fidelity obtained using shares 1 and 2 is strictly greater than any reconstruction involving share 3, it suffices to check the fidelity only for the latter case.

We have seen that whenever player $3$ is involved, the general reconstruction fidelity for a given resource state, using the optimal unity-gain reconstruction protocols discussed in \cref{sec:unitygain}, is
\begin{align}
    \mathcal{F} =     
    \begin{cases}
            2/[3  - \eta^2 + \eta^2 E_{1|2}(g)] & g < 1 \quad(\eta<1) \\
            2/[2 + E_{1|2}(g)] & g = 1 \quad (\eta=1)\\
            2/[3 - 1/\eta^2 + E_{1|2}(g)] & 1 < g < \sqrt{2} \quad(\eta>1)
    \end{cases}
\end{align}
where $\eta(g) = 1/\sqrt{2-g^2}$ and $g \in (0,\sqrt{2})$ is chosen to maximise fidelity. 

Comparing this reconstruction fidelity to the no-cloning limit, $\mathcal{F}>2/3$, we reach our first result defining the entanglement requirements for secure QSS.
\begin{result} \label{result:steering_requirements}
    A sufficient condition for a two-mode resource state to be useful for secure $(2,3)$-threshold QSS with a coherent-state secret is that a $g \in (0,\sqrt{2})$ exists such that the steering parameter satisfies
    \begin{align}
        E_{1|2}(g) < 
        \begin{cases}
            1 & g \le 1, \\
            2 - g^2 & g > 1.
        \end{cases}
        \label{eqn:steering_condition}
    \end{align}
\end{result}

Notably, while this result shows that any resource state exhibiting EPR-steering for some $g\le1$ is useful for secure QSS, a greater magnitude of steering is required when the resource is steerable only for $g>1$. This seeming asymmetry is due to where in the process the amplification correction is implemented. In \textit{lsatt} QSS setups, the secret state is first mixed with the resource mode with both amplified by the QSS scheme before being attenuated afterwards, leaving both secret and resource contributions with no net amplification. However, in \textit{esa} QSS setups, the amplification correction occurs \emph{before} the secret is mixed with the resource, and so the resource contribution is de-amplified by the QSS protocol without a corresponding amplification. The secret state is reproduced with unity gain while the resource contributions are attenuated, reducing their impact on the noise in the output state. Consequently, a higher fidelity can be achieved through \textit{esa}

For a strict implementation of this protocol as described in \cref{sec:qss_overview}, the dealer has access to both resource modes and may choose which mode to mix with the secret state. This free choice of resource mode (i.e., a choice of relabelling modes $1 \leftrightarrow 2$) allows the dealer to decide in which direction this protocol utilises the steering of the resource state, leading to a more general view on the requirements for secure QSS.
\begin{result}
    \label{result:all_steering}
    All EPR-steerable states (one way and two way) are useful for the secure sharing of a coherent-state secret with a suitable dealer allocation of resource modes.
\end{result}
\begin{proof}
    All resource states steerable from mode $2$ to mode $1$ for some $g\le1$ are useful for secure QSS from \cref{result:steering_requirements}. Suppose instead the state is steerable only for $g>1$: such a state is steerable in the opposite direction for $\bar{g} = 1/g < 1$. Hence, this state can be made useful for secure QSS simply by swapping the modes used in the dealer protocol.
\end{proof}
This result requires the dealer to be able to arbitrarily swap resource modes, which we assume is possible in most implementations and discuss further in the Supplemental Material \cite{appendix}. We note that when such swapping is not permitted, \cref{result:all_steering} does not imply that any EPR-steerable state can be used for secure QSS. For example, if distributing one mode of the resource state prior to the other being used in the dealer protocol were desired, one would need to be careful in the choice of resource mode and of any asymmetric degradation of the shares during distribution. 

The lower steering requirement when $g\le1$ also hints at another asymmetry: we show in the supplemental material \cite{appendix} that whenever a resource state is steerable in one direction for some $g>1$, it is always preferable to swap the modes and instead utilise steering in the opposite direction for $\bar{g}=1/g$.

\section{Sharing other Gaussian secrets} \label{sec:other_states}
\begin{figure*}
    \subfloat[][]{
        \includegraphics[width=0.35\linewidth]{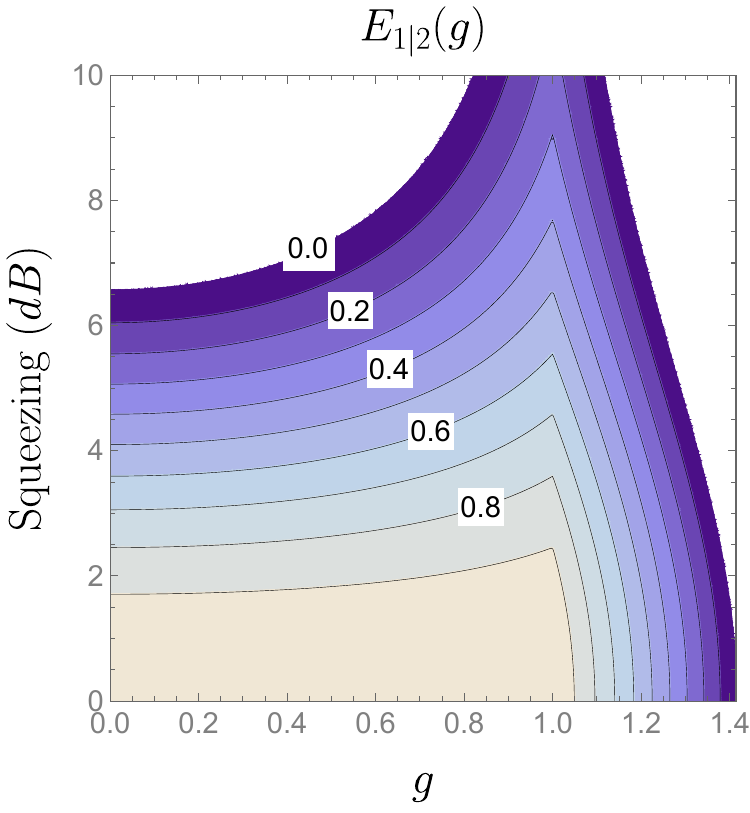}
    }
    \subfloat[][]{
        \includegraphics[width=0.35\linewidth]{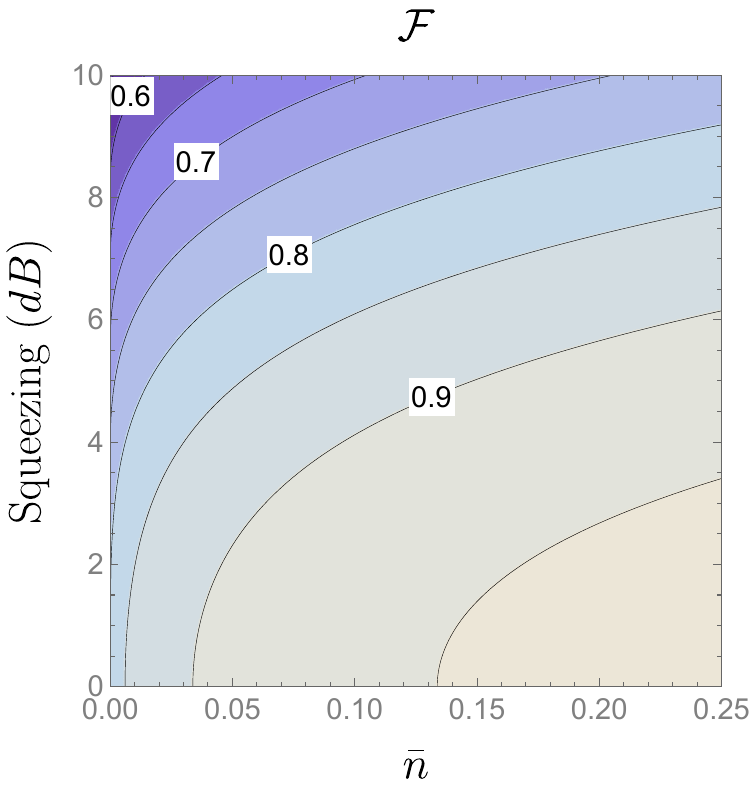}
    }
    \caption{\textbf{(a)} The minimum resource-state steering $\steering$ that guarantees secure QSS for squeezed states. Lower values indicate greater steering, with $\steering=0$ representing perfect steering. Increasing the squeezing parameter increases the resource requirements for secure QSS. There is a clear preference for symmetric resource states, for which the optimal setup is at $g=1$, with greater degrees of steering required to securely share the same squeezed states for asymmetric resources. \textbf{(b)} The fidelity achievable from a QSS setup using a symmetric resource state ($g=1$) with steering parameter $\steering=\frac{1}{2}$. Higher values indicate a more faithful reconstruction, with $\mathcal{F}=1$ representing a perfect copy. Increasing squeezing in the secret state results in reduced reconstruction fidelity, while increasing the thermal photon number allows for a better reconstruction with the same resources.}
    \label{fig:steering}
\end{figure*}

The quantum-state-sharing scheme outlined above generalises naturally to the sharing of any single-mode Gaussian state. In this section we explore the effectiveness of this protocol for some more general classes of Gaussian state: squeezed coherent states and squeezed thermal states.

\subsection{Sharing arbitrary pure Gaussian states} \label{sec:squeezed_coherent}

A squeezed state is a Gaussian state in which the uncertainty in one quadrature has been reduced below the standard quantum limit at the expense of a corresponding increase in the other quadrature. These states have covariance matrix defined by the squeezing parameter $\zeta$, ${\cov = \mathrm{diag}(\e^{-2\zeta},\e^{2\zeta})}$. In general, this squeezing may be along any angle in phase space. As this protocol is phase-independent, however, we may neglect the squeezing angle and so assume for convenience that the states are squeezed along the $\qx$ and $\qp$ quadratures.

We consider in this section the use of our QSS protocol for the sharing of squeezed coherent states - the most general pure Gaussian state. We show in the Supplemental Material \cite{appendix} that after QSS these states can be reconstructed with fidelity
\begin{align}
    \mathcal{F} &= \frac{2}{\sqrt{(2\e^{2\zeta}+\chi)(2\e^{-2\zeta}+\chi)}},
    \label{eqn:pure_esa_fidelity}
\end{align}
where 
\begin{align}
    \chi = \begin{cases}
        \eta^2 \steering + 1 - \eta^2 & g \le 1 \quad \textit{(esa)} \\
        \steering + 1 - \frac{1}{\eta^2} & g \ge 1 \quad \textit{(lsatt)}
    \end{cases}
\end{align}
represents the $g$-dependent component introduced by the amplification correction. Increasing the squeezing $\zeta$ in the secret state reduces the achievable reconstruction fidelity. This fidelity for squeezed Gaussian states is shown along the $\bar{n}=0$ axis in \cref{fig:steering}(b). 

We now turn to the question of security. It is in-general more difficult to clone states with an unknown squeezing than coherent states, with strategies optimal for cloning coherent states unable to achieve $\mathcal{F}=2/3$ cloning fidelity when applied to states with unknown squeezing \cite{olivares_2006}. The optimal protocol for cloning squeezed states is not known, and so reaching the $\mathcal{F}>2/3$ bound may not be necessary for security. However, the cloning fidelity for squeezed states remains upper bounded by $\mathcal{F}=2/3$ and so this condition is still sufficient for security \cite{cerf_2000a}. In the absence of an optimal protocol, here we use this upper bound as our threshold for guaranteed security.

From this fidelity threshold we can derive the following sufficient condition for the protocol's security.

\begin{result}
    A QSS protocol for the sharing of a pure Gaussian secret state with squeezing of up to $\zeta_\text{max}$ is secure if the resource state used has steering of
    \begin{align}
        \steering < 
        \begin{cases}
            1 - \frac{1}{\eta^2} \Gamma(\zeta_\text{max}) & g \le 1 \quad(\eta\le1) \\
            \frac{1}{\eta^2} - \Gamma(\zeta_\text{max}) & g \ge 1 \quad(\eta\ge1)
        \end{cases},
    \end{align}
    for some $g\in(0,\sqrt{2})$ where $1/\eta^2 = 2-g^2$ and
    \begin{align}
        \Gamma(\zeta) = 1 + 2\cosh(2\zeta) - \sqrt{4\cosh^2(2\zeta)+5} \ge 0
    \end{align}
    is a monotonically increasing function of $\zeta$ with $\Gamma(0)=0$.
\end{result}

This result is shown in \cref{fig:steering}(a). Comparing this condition to that for coherent states in \cref{eqn:steering_condition}, the effect of squeezing the secret state on QSS becomes apparent. Securely sharing a secret state with one quadrature squeezed below the vacuum limit requires a corresponding increase in entanglement above what is necessary for coherent states. The preference for symmetric resources, for which $g=1$ remains, with less-entangled resources capable of securely sharing more squeezing when utilised symmetrically. Notably, the required extra steering tends to $\Gamma=1$ as $\zeta\rightarrow\infty$, so even highly squeezed states and, in the limit, quadrature states can be shared securely with a suitably entangled resource state.

\subsection{Sharing arbitrary mixed Gaussian states} 
\label{sec:squeezed_thermal}

Finally, we briefly discuss the potential use of this protocol for squeezed displaced thermal states: the most general possible single-mode Gaussian state. These thermal states have covariance matrix $\cov=\tilde{n}\,\mathrm{diag}(\exp(-2\zeta),\exp(2\zeta))$, where $\tilde{n} = (2\bar{n}+1)$ represents the average number of thermal photons $\bar{n}$ in the state prior to displacement and $\zeta$ again represents the degree to which the state is squeezed.

We show in the supplemental material \cite{appendix} that the fidelity when sharing such states using a given resource state increases with increasing thermal photon number $\bar{n}$ and decreases with increasing squeezing $\zeta$. As before, it does not depend on the mean amplitude $\mean$. Utilising the appropriate amplification correction after the QSS stage, this protocol can achieve a reconstruction fidelity of
\begin{widetext}
    \begin{align}
        \mathcal{F} = \frac{2}{ \sqrt{\left( \tilde{n} \chi + (\tilde{n}^2 + 1) \e^{2\zeta} \right) \left( \tilde{n} \chi + (\tilde{n}^2 + 1) \e^{-2\zeta} \right)} - \sqrt{(\tilde{n}^2 - 1) \left( \tilde{n}^2 + \chi^2 + 2 \tilde{n} \chi \cosh(2\zeta) - 1 \right)}},
    \end{align}
\end{widetext}
where $\chi$ represents the impact of the amplification correction as before and is dependent on the value of $g$. This represents the most general measure of ideal reconstruction fidelity for the sharing of any single-mode Gaussian state. The impact of both squeezing and average thermal photon number on the reconstruction fidelity is shown in \cref{fig:steering} (b).

Once one considers states with added thermal noise above the uncertainty limit, a greater cloning fidelity is achievable and so reaching the $\mathcal{F}>2/3$ threshold is no longer sufficient for security. Consequently we do not present a condition on the security of this scheme for thermal states. Some work has been done on the question of cloning thermal states \cite{olivares_2006,guta_2006}, but optimality has not yet been shown for cloning fidelity. The use of this scheme for thermal states should then be carefully considered as security may not be guaranteed.

\section{Conclusion}
We have shown here that the secure sharing of coherent states between three players is possible using a resource state exhibiting any form of EPR steering. Notably, this is a looser requirement than those for secure quantum teleportation, where a resource state must be two-way steerable to be useful \cite{he_2015a}. Consequently, QSS could be considered a competitive alternative to quantum teleportation for secure state distribution when a set of communication channels can be trusted only collectively.

Going beyond coherent states, we have analysed this QSS protocol for any single-mode Gaussian state, including squeezed states and thermal states. We have shown that while increased resource steering is required to share squeezed states, any pure single-mode Gaussian state is securely sharable with a suitably entangled resource state. Future work on this subject could involve a generalisation to securely share the wider class of multi-mode Gaussian secret states. In such a case it will be of critical importance to preserve correlations between modes of the secret state. Additionally, the practical implementations of the analogous $(k,n)$-threshold QSS should be considered.

There is potential for QSS schemes to find uses in blind quantum computation \cite{ouyang_2017} or as quantum Reed-Solomon codes for error correction \cite{grassl_1999}. By splitting the original state into shares and transmitting the shares separately, the original state can be reconstructed even should some of the shares lose fidelity in the transmission. Quantum error correction is an important sub-routine both in quantum computing and in quantum state distribution. QSS schemes then may contribute to the practical implementation of a future quantum internet \cite{wehner_2018}.

\acknowledgments
This work has been supported by the Scottish Universities Physics Alliance (SUPA) and by the Engineering and Physical Sciences Research Council (EPSRC). C.W. acknowledges the funding from the Global Ph.D. program of the University of St. Andrews and Macquarie University. M.T. and N.K. were supported by the EU Flagship on Quantum Technologies, project PhoG (820365).

\bibliography{bibliography.bib,supmat.bib}

\begin{thebibliography}{29}%
\makeatletter
\providecommand \@ifxundefined [1]{%
 \@ifx{#1\undefined}
}%
\providecommand \@ifnum [1]{%
 \ifnum #1\expandafter \@firstoftwo
 \else \expandafter \@secondoftwo
 \fi
}%
\providecommand \@ifx [1]{%
 \ifx #1\expandafter \@firstoftwo
 \else \expandafter \@secondoftwo
 \fi
}%
\providecommand \natexlab [1]{#1}%
\providecommand \enquote  [1]{``#1''}%
\providecommand \bibnamefont  [1]{#1}%
\providecommand \bibfnamefont [1]{#1}%
\providecommand \citenamefont [1]{#1}%
\providecommand \href@noop [0]{\@secondoftwo}%
\providecommand \href [0]{\begingroup \@sanitize@url \@href}%
\providecommand \@href[1]{\@@startlink{#1}\@@href}%
\providecommand \@@href[1]{\endgroup#1\@@endlink}%
\providecommand \@sanitize@url [0]{\catcode `\\12\catcode `\$12\catcode
  `\&12\catcode `\#12\catcode `\^12\catcode `\_12\catcode `\%12\relax}%
\providecommand \@@startlink[1]{}%
\providecommand \@@endlink[0]{}%
\providecommand \url  [0]{\begingroup\@sanitize@url \@url }%
\providecommand \@url [1]{\endgroup\@href {#1}{\urlprefix }}%
\providecommand \urlprefix  [0]{URL }%
\providecommand \Eprint [0]{\href }%
\providecommand \doibase [0]{https://doi.org/}%
\providecommand \selectlanguage [0]{\@gobble}%
\providecommand \bibinfo  [0]{\@secondoftwo}%
\providecommand \bibfield  [0]{\@secondoftwo}%
\providecommand \translation [1]{[#1]}%
\providecommand \BibitemOpen [0]{}%
\providecommand \bibitemStop [0]{}%
\providecommand \bibitemNoStop [0]{.\EOS\space}%
\providecommand \EOS [0]{\spacefactor3000\relax}%
\providecommand \BibitemShut  [1]{\csname bibitem#1\endcsname}%
\let\auto@bib@innerbib\@empty
\bibitem [{\citenamefont {Shamir}(1979)}]{shamir_1979}%
  \BibitemOpen
  \bibfield  {author} {\bibinfo {author} {\bibfnamefont {A.}~\bibnamefont
  {Shamir}},\ }\href {https://doi.org/10.1145/359168.359176} {\bibfield
  {journal} {\bibinfo  {journal} {Communications of the ACM}\ }\textbf
  {\bibinfo {volume} {22}},\ \bibinfo {pages} {612} (\bibinfo {year}
  {1979})}\BibitemShut {NoStop}%
\bibitem [{\citenamefont {Blakley}(1979)}]{blakley_1979}%
  \BibitemOpen
  \bibfield  {author} {\bibinfo {author} {\bibfnamefont {G.~R.}\ \bibnamefont
  {Blakley}},\ }in\ \href {https://doi.org/10.1109/MARK.1979.8817296} {\emph
  {\bibinfo {booktitle} {1979 {{International Workshop}} on {{Managing
  Requirements Knowledge}} ({{MARK}})}}}\ (\bibinfo  {publisher} {{IEEE}},\
  \bibinfo {address} {{New York, NY, USA}},\ \bibinfo {year} {1979})\ pp.\
  \bibinfo {pages} {313--318}\BibitemShut {NoStop}%
\bibitem [{\citenamefont {Schneier}(1996)}]{schneier_1996}%
  \BibitemOpen
  \bibfield  {author} {\bibinfo {author} {\bibfnamefont {B.}~\bibnamefont
  {Schneier}},\ }\href@noop {} {\emph {\bibinfo {title} {Applied Cryptography:
  Protocols, Algorithms, and Source Code in {{C}}}}},\ \bibinfo {edition}
  {2nd}\ ed.\ (\bibinfo  {publisher} {{Wiley}},\ \bibinfo {address} {{New
  York}},\ \bibinfo {year} {1996})\BibitemShut {NoStop}%
\bibitem [{Note1()}]{Note1}%
  \BibitemOpen
  \bibinfo {note} {As distinct from quantum secret sharing, which uses quantum
  resources to securely share classical information \cite
  {richter_2021}.}\BibitemShut {Stop}%
\bibitem [{\citenamefont {Cleve}\ \emph {et~al.}(1999)\citenamefont {Cleve},
  \citenamefont {Gottesman},\ and\ \citenamefont {Lo}}]{cleve_1999}%
  \BibitemOpen
  \bibfield  {author} {\bibinfo {author} {\bibfnamefont {R.}~\bibnamefont
  {Cleve}}, \bibinfo {author} {\bibfnamefont {D.}~\bibnamefont {Gottesman}},\
  and\ \bibinfo {author} {\bibfnamefont {H.-K.}\ \bibnamefont {Lo}},\ }\href
  {https://doi.org/10.1103/PhysRevLett.83.648} {\bibfield  {journal} {\bibinfo
  {journal} {Physical Review Letters}\ }\textbf {\bibinfo {volume} {83}},\
  \bibinfo {pages} {648} (\bibinfo {year} {1999})}\BibitemShut {NoStop}%
\bibitem [{\citenamefont {Hillery}\ \emph {et~al.}(1999)\citenamefont
  {Hillery}, \citenamefont {Bu{\v z}ek},\ and\ \citenamefont
  {Berthiaume}}]{hillery_1999}%
  \BibitemOpen
  \bibfield  {author} {\bibinfo {author} {\bibfnamefont {M.}~\bibnamefont
  {Hillery}}, \bibinfo {author} {\bibfnamefont {V.}~\bibnamefont {Bu{\v
  z}ek}},\ and\ \bibinfo {author} {\bibfnamefont {A.}~\bibnamefont
  {Berthiaume}},\ }\href {https://doi.org/10.1103/PhysRevA.59.1829} {\bibfield
  {journal} {\bibinfo  {journal} {Physical Review A}\ }\textbf {\bibinfo
  {volume} {59}},\ \bibinfo {pages} {1829} (\bibinfo {year}
  {1999})}\BibitemShut {NoStop}%
\bibitem [{\citenamefont {Ouyang}\ \emph {et~al.}(2017)\citenamefont {Ouyang},
  \citenamefont {Tan}, \citenamefont {Zhao},\ and\ \citenamefont
  {Fitzsimons}}]{ouyang_2017}%
  \BibitemOpen
  \bibfield  {author} {\bibinfo {author} {\bibfnamefont {Y.}~\bibnamefont
  {Ouyang}}, \bibinfo {author} {\bibfnamefont {S.-H.}\ \bibnamefont {Tan}},
  \bibinfo {author} {\bibfnamefont {L.}~\bibnamefont {Zhao}},\ and\ \bibinfo
  {author} {\bibfnamefont {J.~F.}\ \bibnamefont {Fitzsimons}},\ }\href
  {https://doi.org/10.1103/PhysRevA.96.052333} {\bibfield  {journal} {\bibinfo
  {journal} {Physical Review A}\ }\textbf {\bibinfo {volume} {96}},\ \bibinfo
  {pages} {052333} (\bibinfo {year} {2017})}\BibitemShut {NoStop}%
\bibitem [{\citenamefont {Grassl}\ \emph {et~al.}(1999)\citenamefont {Grassl},
  \citenamefont {Geiselmann},\ and\ \citenamefont {Beth}}]{grassl_1999}%
  \BibitemOpen
  \bibfield  {author} {\bibinfo {author} {\bibfnamefont {M.}~\bibnamefont
  {Grassl}}, \bibinfo {author} {\bibfnamefont {W.}~\bibnamefont {Geiselmann}},\
  and\ \bibinfo {author} {\bibfnamefont {T.}~\bibnamefont {Beth}},\ }in\ \href
  {https://doi.org/10.1007/3-540-46796-3_23} {\emph {\bibinfo {booktitle}
  {Applied {{Algebra}}, {{Algebraic Algorithms}} and {{Error-Correcting
  Codes}}}}},\ \bibinfo {series} {Lecture {{Notes}} in {{Computer Science}}},
  Vol.\ \bibinfo {volume} {1719},\ \bibinfo {editor} {edited by\ \bibinfo
  {editor} {\bibfnamefont {M.}~\bibnamefont {Fossorier}}, \bibinfo {editor}
  {\bibfnamefont {H.}~\bibnamefont {Imai}}, \bibinfo {editor} {\bibfnamefont
  {S.}~\bibnamefont {Lin}},\ and\ \bibinfo {editor} {\bibfnamefont
  {A.}~\bibnamefont {Poli}}}\ (\bibinfo  {publisher} {{Springer}},\ \bibinfo
  {address} {{Berlin, Heidelberg}},\ \bibinfo {year} {1999})\ pp.\ \bibinfo
  {pages} {231--244}\BibitemShut {NoStop}%
\bibitem [{\citenamefont {Wehner}\ \emph {et~al.}(2018)\citenamefont {Wehner},
  \citenamefont {Elkouss},\ and\ \citenamefont {Hanson}}]{wehner_2018}%
  \BibitemOpen
  \bibfield  {author} {\bibinfo {author} {\bibfnamefont {S.}~\bibnamefont
  {Wehner}}, \bibinfo {author} {\bibfnamefont {D.}~\bibnamefont {Elkouss}},\
  and\ \bibinfo {author} {\bibfnamefont {R.}~\bibnamefont {Hanson}},\ }\href
  {https://doi.org/10.1126/science.aam9288} {\bibfield  {journal} {\bibinfo
  {journal} {Science}\ }\textbf {\bibinfo {volume} {362}},\ \bibinfo {pages}
  {eaam9288} (\bibinfo {year} {2018})}\BibitemShut {NoStop}%
\bibitem [{\citenamefont {Ball}\ \emph {et~al.}(2021)\citenamefont {Ball},
  \citenamefont {Biercuk},\ and\ \citenamefont {Hush}}]{ball_2021}%
  \BibitemOpen
  \bibfield  {author} {\bibinfo {author} {\bibfnamefont {H.}~\bibnamefont
  {Ball}}, \bibinfo {author} {\bibfnamefont {M.~J.}\ \bibnamefont {Biercuk}},\
  and\ \bibinfo {author} {\bibfnamefont {M.~R.}\ \bibnamefont {Hush}},\ }\href
  {https://doi.org/10.1063/PT.3.4698} {\bibfield  {journal} {\bibinfo
  {journal} {Physics Today}\ }\textbf {\bibinfo {volume} {74}},\ \bibinfo
  {pages} {28} (\bibinfo {year} {2021})}\BibitemShut {NoStop}%
\bibitem [{\citenamefont {Tyc}\ and\ \citenamefont {Sanders}(2002)}]{tyc_2002}%
  \BibitemOpen
  \bibfield  {author} {\bibinfo {author} {\bibfnamefont {T.}~\bibnamefont
  {Tyc}}\ and\ \bibinfo {author} {\bibfnamefont {B.~C.}\ \bibnamefont
  {Sanders}},\ }\href {https://doi.org/10.1103/PhysRevA.65.042310} {\bibfield
  {journal} {\bibinfo  {journal} {Physical Review A}\ }\textbf {\bibinfo
  {volume} {65}},\ \bibinfo {pages} {042310} (\bibinfo {year}
  {2002})}\BibitemShut {NoStop}%
\bibitem [{\citenamefont {Lance}\ \emph {et~al.}(2003)\citenamefont {Lance},
  \citenamefont {Symul}, \citenamefont {Bowen}, \citenamefont {Tyc},
  \citenamefont {Sanders},\ and\ \citenamefont {Lam}}]{lance_2003}%
  \BibitemOpen
  \bibfield  {author} {\bibinfo {author} {\bibfnamefont {A.~M.}\ \bibnamefont
  {Lance}}, \bibinfo {author} {\bibfnamefont {T.}~\bibnamefont {Symul}},
  \bibinfo {author} {\bibfnamefont {W.~P.}\ \bibnamefont {Bowen}}, \bibinfo
  {author} {\bibfnamefont {T.}~\bibnamefont {Tyc}}, \bibinfo {author}
  {\bibfnamefont {B.~C.}\ \bibnamefont {Sanders}},\ and\ \bibinfo {author}
  {\bibfnamefont {P.~K.}\ \bibnamefont {Lam}},\ }\bibfield  {journal} {\bibinfo
   {journal} {New Journal of Physics}\ }\textbf {\bibinfo {volume} {5}},\ \href
  {https://doi.org/10.1088/1367-2630/5/1/304} {10.1088/1367-2630/5/1/304}
  (\bibinfo {year} {2003})\BibitemShut {NoStop}%
\bibitem [{\citenamefont {Lance}\ \emph {et~al.}(2004)\citenamefont {Lance},
  \citenamefont {Symul}, \citenamefont {Bowen}, \citenamefont {Sanders},\ and\
  \citenamefont {Lam}}]{lance_2004}%
  \BibitemOpen
  \bibfield  {author} {\bibinfo {author} {\bibfnamefont {A.~M.}\ \bibnamefont
  {Lance}}, \bibinfo {author} {\bibfnamefont {T.}~\bibnamefont {Symul}},
  \bibinfo {author} {\bibfnamefont {W.~P.}\ \bibnamefont {Bowen}}, \bibinfo
  {author} {\bibfnamefont {B.~C.}\ \bibnamefont {Sanders}},\ and\ \bibinfo
  {author} {\bibfnamefont {P.~K.}\ \bibnamefont {Lam}},\ }\href
  {https://doi.org/10.1103/PhysRevLett.92.177903} {\bibfield  {journal}
  {\bibinfo  {journal} {Physical Review Letters}\ }\textbf {\bibinfo {volume}
  {92}},\ \bibinfo {pages} {177903} (\bibinfo {year} {2004})}\BibitemShut
  {NoStop}%
\bibitem [{\citenamefont {Lance}\ \emph {et~al.}(2005)\citenamefont {Lance},
  \citenamefont {Symul}, \citenamefont {Bowen}, \citenamefont {Sanders},
  \citenamefont {Tyc}, \citenamefont {Ralph},\ and\ \citenamefont
  {Lam}}]{lance_2005}%
  \BibitemOpen
  \bibfield  {author} {\bibinfo {author} {\bibfnamefont {A.~M.}\ \bibnamefont
  {Lance}}, \bibinfo {author} {\bibfnamefont {T.}~\bibnamefont {Symul}},
  \bibinfo {author} {\bibfnamefont {W.~P.}\ \bibnamefont {Bowen}}, \bibinfo
  {author} {\bibfnamefont {B.~C.}\ \bibnamefont {Sanders}}, \bibinfo {author}
  {\bibfnamefont {T.}~\bibnamefont {Tyc}}, \bibinfo {author} {\bibfnamefont
  {T.~C.}\ \bibnamefont {Ralph}},\ and\ \bibinfo {author} {\bibfnamefont
  {P.~K.}\ \bibnamefont {Lam}},\ }\href
  {https://doi.org/10.1103/PhysRevA.71.033814} {\bibfield  {journal} {\bibinfo
  {journal} {Physical Review A}\ }\textbf {\bibinfo {volume} {71}},\ \bibinfo
  {pages} {033814} (\bibinfo {year} {2005})}\BibitemShut {NoStop}%
\bibitem [{\citenamefont {Chitambar}\ and\ \citenamefont
  {Gour}(2019)}]{chitambar_2019}%
  \BibitemOpen
  \bibfield  {author} {\bibinfo {author} {\bibfnamefont {E.}~\bibnamefont
  {Chitambar}}\ and\ \bibinfo {author} {\bibfnamefont {G.}~\bibnamefont
  {Gour}},\ }\href {https://doi.org/10.1103/RevModPhys.91.025001} {\bibfield
  {journal} {\bibinfo  {journal} {Reviews of Modern Physics}\ }\textbf
  {\bibinfo {volume} {91}},\ \bibinfo {pages} {025001} (\bibinfo {year}
  {2019})},\ \Eprint {https://arxiv.org/abs/1806.06107} {arxiv:1806.06107}
  \BibitemShut {NoStop}%
\bibitem [{\citenamefont {Weedbrook}\ \emph {et~al.}(2012)\citenamefont
  {Weedbrook}, \citenamefont {Pirandola}, \citenamefont
  {{Garc{\'i}a-Patr{\'o}n}}, \citenamefont {Cerf}, \citenamefont {Ralph},
  \citenamefont {Shapiro},\ and\ \citenamefont {Lloyd}}]{weedbrook_2012}%
  \BibitemOpen
  \bibfield  {author} {\bibinfo {author} {\bibfnamefont {C.}~\bibnamefont
  {Weedbrook}}, \bibinfo {author} {\bibfnamefont {S.}~\bibnamefont
  {Pirandola}}, \bibinfo {author} {\bibfnamefont {R.}~\bibnamefont
  {{Garc{\'i}a-Patr{\'o}n}}}, \bibinfo {author} {\bibfnamefont {N.~J.}\
  \bibnamefont {Cerf}}, \bibinfo {author} {\bibfnamefont {T.~C.}\ \bibnamefont
  {Ralph}}, \bibinfo {author} {\bibfnamefont {J.~H.}\ \bibnamefont {Shapiro}},\
  and\ \bibinfo {author} {\bibfnamefont {S.}~\bibnamefont {Lloyd}},\ }\href
  {https://doi.org/10.1103/RevModPhys.84.621} {\bibfield  {journal} {\bibinfo
  {journal} {Reviews of Modern Physics}\ }\textbf {\bibinfo {volume} {84}},\
  \bibinfo {pages} {621} (\bibinfo {year} {2012})}\BibitemShut {NoStop}%
\bibitem [{\citenamefont {Wiseman}\ \emph {et~al.}(2007)\citenamefont
  {Wiseman}, \citenamefont {Jones},\ and\ \citenamefont
  {Doherty}}]{wiseman_2007}%
  \BibitemOpen
  \bibfield  {author} {\bibinfo {author} {\bibfnamefont {H.~M.}\ \bibnamefont
  {Wiseman}}, \bibinfo {author} {\bibfnamefont {S.~J.}\ \bibnamefont {Jones}},\
  and\ \bibinfo {author} {\bibfnamefont {A.~C.}\ \bibnamefont {Doherty}},\
  }\href {https://doi.org/10.1103/PhysRevLett.98.140402} {\bibfield  {journal}
  {\bibinfo  {journal} {Physical Review Letters}\ }\textbf {\bibinfo {volume}
  {98}},\ \bibinfo {pages} {140402} (\bibinfo {year} {2007})}\BibitemShut
  {NoStop}%
\bibitem [{\citenamefont {Reid}(1989)}]{reid_1989}%
  \BibitemOpen
  \bibfield  {author} {\bibinfo {author} {\bibfnamefont {M.~D.}\ \bibnamefont
  {Reid}},\ }\href {https://doi.org/10.1103/PhysRevA.40.913} {\bibfield
  {journal} {\bibinfo  {journal} {Physical Review A}\ }\textbf {\bibinfo
  {volume} {40}},\ \bibinfo {pages} {913} (\bibinfo {year} {1989})}\BibitemShut
  {NoStop}%
\bibitem [{\citenamefont {He}\ \emph {et~al.}(2015{\natexlab{a}})\citenamefont
  {He}, \citenamefont {Gong},\ and\ \citenamefont {Reid}}]{he_2015a}%
  \BibitemOpen
  \bibfield  {author} {\bibinfo {author} {\bibfnamefont {Q.~Y.}\ \bibnamefont
  {He}}, \bibinfo {author} {\bibfnamefont {Q.~H.}\ \bibnamefont {Gong}},\ and\
  \bibinfo {author} {\bibfnamefont {M.~D.}\ \bibnamefont {Reid}},\ }\href
  {https://doi.org/10.1103/PhysRevLett.114.060402} {\bibfield  {journal}
  {\bibinfo  {journal} {Physical Review Letters}\ }\textbf {\bibinfo {volume}
  {114}},\ \bibinfo {pages} {060402} (\bibinfo {year}
  {2015}{\natexlab{a}})}\BibitemShut {NoStop}%
\bibitem [{Note2()}]{Note2}%
  \BibitemOpen
  \bibinfo {note} {This can be seen by considering the channel resulting in
  $\protect \hat {X}^\pm _\protect \text {out}=1/\protect \sqrt
  {1-g^2}(\protect \hat {X}^\pm _{r1} \mp g\protect \hat {X}^\pm _{r2})$ for
  $g<1$ and imposing the uncertainty limit on this output to get a condition
  for minimum $E_{1|2}(g)>1-g^2$. This can be converted to a condition on
  steering in the opposite direction for $\protect \bar {g}=1/g>1$ of
  $E_{2|1}(\protect \bar {g})>\protect \bar {g}^2-1$ and so steering can only
  be certified ($E_{1|2}(g)<1$) for values of $g\in (0,\protect \sqrt
  {2})$.}\BibitemShut {Stop}%
\bibitem [{app()}]{appendix}%
  \BibitemOpen
  \href@noop {} {}\bibinfo {note} {Supplementary Material}\BibitemShut
  {NoStop}%
\bibitem [{\citenamefont {He}\ \emph {et~al.}(2015{\natexlab{b}})\citenamefont
  {He}, \citenamefont {{Rosales-Z{\'a}rate}}, \citenamefont {Adesso},\ and\
  \citenamefont {Reid}}]{he_2015}%
  \BibitemOpen
  \bibfield  {author} {\bibinfo {author} {\bibfnamefont {Q.}~\bibnamefont
  {He}}, \bibinfo {author} {\bibfnamefont {L.}~\bibnamefont
  {{Rosales-Z{\'a}rate}}}, \bibinfo {author} {\bibfnamefont {G.}~\bibnamefont
  {Adesso}},\ and\ \bibinfo {author} {\bibfnamefont {M.~D.}\ \bibnamefont
  {Reid}},\ }\href {https://doi.org/10.1103/PhysRevLett.115.180502} {\bibfield
  {journal} {\bibinfo  {journal} {Physical Review Letters}\ }\textbf {\bibinfo
  {volume} {115}},\ \bibinfo {pages} {180502} (\bibinfo {year}
  {2015}{\natexlab{b}})}\BibitemShut {NoStop}%
\bibitem [{\citenamefont {Grosshans}\ and\ \citenamefont
  {Grangier}(2001)}]{grosshans_2001}%
  \BibitemOpen
  \bibfield  {author} {\bibinfo {author} {\bibfnamefont {F.}~\bibnamefont
  {Grosshans}}\ and\ \bibinfo {author} {\bibfnamefont {P.}~\bibnamefont
  {Grangier}},\ }\href {https://doi.org/10.1103/PhysRevA.64.010301} {\bibfield
  {journal} {\bibinfo  {journal} {Physical Review A}\ }\textbf {\bibinfo
  {volume} {64}},\ \bibinfo {pages} {010301(R)} (\bibinfo {year}
  {2001})}\BibitemShut {NoStop}%
\bibitem [{\citenamefont {Cerf}\ and\ \citenamefont
  {Iblisdir}(2000)}]{cerf_2000}%
  \BibitemOpen
  \bibfield  {author} {\bibinfo {author} {\bibfnamefont {N.~J.}\ \bibnamefont
  {Cerf}}\ and\ \bibinfo {author} {\bibfnamefont {S.}~\bibnamefont
  {Iblisdir}},\ }\href {https://doi.org/10.1103/PhysRevA.62.040301} {\bibfield
  {journal} {\bibinfo  {journal} {Physical Review A}\ }\textbf {\bibinfo
  {volume} {62}},\ \bibinfo {pages} {040301} (\bibinfo {year}
  {2000})}\BibitemShut {NoStop}%
\bibitem [{\citenamefont {Cerf}\ \emph {et~al.}(2005)\citenamefont {Cerf},
  \citenamefont {Kr{\"u}ger}, \citenamefont {Navez}, \citenamefont {Werner},\
  and\ \citenamefont {Wolf}}]{cerf_2005}%
  \BibitemOpen
  \bibfield  {author} {\bibinfo {author} {\bibfnamefont {N.~J.}\ \bibnamefont
  {Cerf}}, \bibinfo {author} {\bibfnamefont {O.}~\bibnamefont {Kr{\"u}ger}},
  \bibinfo {author} {\bibfnamefont {P.}~\bibnamefont {Navez}}, \bibinfo
  {author} {\bibfnamefont {R.~F.}\ \bibnamefont {Werner}},\ and\ \bibinfo
  {author} {\bibfnamefont {M.~M.}\ \bibnamefont {Wolf}},\ }\href
  {https://doi.org/10.1103/PhysRevLett.95.070501} {\bibfield  {journal}
  {\bibinfo  {journal} {Physical Review Letters}\ }\textbf {\bibinfo {volume}
  {95}},\ \bibinfo {pages} {070501} (\bibinfo {year} {2005})}\BibitemShut
  {NoStop}%
\bibitem [{\citenamefont {Olivares}\ \emph {et~al.}(2006)\citenamefont
  {Olivares}, \citenamefont {Paris},\ and\ \citenamefont
  {Andersen}}]{olivares_2006}%
  \BibitemOpen
  \bibfield  {author} {\bibinfo {author} {\bibfnamefont {S.}~\bibnamefont
  {Olivares}}, \bibinfo {author} {\bibfnamefont {M.~G.~A.}\ \bibnamefont
  {Paris}},\ and\ \bibinfo {author} {\bibfnamefont {U.~L.}\ \bibnamefont
  {Andersen}},\ }\href {https://doi.org/10.1103/PhysRevA.73.062330} {\bibfield
  {journal} {\bibinfo  {journal} {Physical Review A}\ }\textbf {\bibinfo
  {volume} {73}},\ \bibinfo {pages} {062330} (\bibinfo {year}
  {2006})}\BibitemShut {NoStop}%
\bibitem [{\citenamefont {Cerf}\ \emph {et~al.}(2000)\citenamefont {Cerf},
  \citenamefont {Ipe},\ and\ \citenamefont {Rottenberg}}]{cerf_2000a}%
  \BibitemOpen
  \bibfield  {author} {\bibinfo {author} {\bibfnamefont {N.~J.}\ \bibnamefont
  {Cerf}}, \bibinfo {author} {\bibfnamefont {A.}~\bibnamefont {Ipe}},\ and\
  \bibinfo {author} {\bibfnamefont {X.}~\bibnamefont {Rottenberg}},\ }\href
  {https://doi.org/10.1103/PhysRevLett.85.1754} {\bibfield  {journal} {\bibinfo
   {journal} {Physical Review Letters}\ }\textbf {\bibinfo {volume} {85}},\
  \bibinfo {pages} {1754} (\bibinfo {year} {2000})}\BibitemShut {NoStop}%
\bibitem [{\citenamefont {Gu{\c t}{\u a}}\ and\ \citenamefont
  {Matsumoto}(2006)}]{guta_2006}%
  \BibitemOpen
  \bibfield  {author} {\bibinfo {author} {\bibfnamefont {M.}~\bibnamefont
  {Gu{\c t}{\u a}}}\ and\ \bibinfo {author} {\bibfnamefont {K.}~\bibnamefont
  {Matsumoto}},\ }\href {https://doi.org/10.1103/PhysRevA.74.032305} {\bibfield
   {journal} {\bibinfo  {journal} {Physical Review A}\ }\textbf {\bibinfo
  {volume} {74}},\ \bibinfo {pages} {032305} (\bibinfo {year}
  {2006})}\BibitemShut {NoStop}%
\bibitem [{\citenamefont {Richter}\ \emph {et~al.}(2021)\citenamefont
  {Richter}, \citenamefont {Thornton}, \citenamefont {Khan}, \citenamefont
  {Scott}, \citenamefont {Jaksch}, \citenamefont {Vogl}, \citenamefont
  {Stiller}, \citenamefont {Leuchs}, \citenamefont {Marquardt},\ and\
  \citenamefont {Korolkova}}]{richter_2021}%
  \BibitemOpen
  \bibfield  {author} {\bibinfo {author} {\bibfnamefont {S.}~\bibnamefont
  {Richter}}, \bibinfo {author} {\bibfnamefont {M.}~\bibnamefont {Thornton}},
  \bibinfo {author} {\bibfnamefont {I.}~\bibnamefont {Khan}}, \bibinfo {author}
  {\bibfnamefont {H.}~\bibnamefont {Scott}}, \bibinfo {author} {\bibfnamefont
  {K.}~\bibnamefont {Jaksch}}, \bibinfo {author} {\bibfnamefont
  {U.}~\bibnamefont {Vogl}}, \bibinfo {author} {\bibfnamefont {B.}~\bibnamefont
  {Stiller}}, \bibinfo {author} {\bibfnamefont {G.}~\bibnamefont {Leuchs}},
  \bibinfo {author} {\bibfnamefont {C.}~\bibnamefont {Marquardt}},\ and\
  \bibinfo {author} {\bibfnamefont {N.}~\bibnamefont {Korolkova}},\ }\href
  {https://doi.org/10.1103/PhysRevX.11.011038} {\bibfield  {journal} {\bibinfo
  {journal} {Physical Review X}\ }\textbf {\bibinfo {volume} {11}},\ \bibinfo
  {pages} {011038} (\bibinfo {year} {2021})}\BibitemShut {NoStop}%
\end{thebibliography}%


\begin{thebibliography}{5}
\expandafter\ifx\csname natexlab\endcsname\relax\def\natexlab#1{#1}\fi
\expandafter\ifx\csname bibnamefont\endcsname\relax
  \def\bibnamefont#1{#1}\fi
\expandafter\ifx\csname bibfnamefont\endcsname\relax
  \def\bibfnamefont#1{#1}\fi
\expandafter\ifx\csname citenamefont\endcsname\relax
  \def\citenamefont#1{#1}\fi
\expandafter\ifx\csname url\endcsname\relax
  \def\url#1{\texttt{#1}}\fi
\expandafter\ifx\csname urlprefix\endcsname\relax\def\urlprefix{URL }\fi
\providecommand{\bibinfo}[2]{#2}
\providecommand{\eprint}[2][]{\url{#2}}

\bibitem[{\citenamefont{Weedbrook et~al.}(2012)\citenamefont{Weedbrook,
  Pirandola, {Garc{\'i}a-Patr{\'o}n}, Cerf, Ralph, Shapiro, and
  Lloyd}}]{weedbrook_2012}
\bibinfo{author}{\bibfnamefont{C.}~\bibnamefont{Weedbrook}},
  \bibinfo{author}{\bibfnamefont{S.}~\bibnamefont{Pirandola}},
  \bibinfo{author}{\bibfnamefont{R.}~\bibnamefont{{Garc{\'i}a-Patr{\'o}n}}},
  \bibinfo{author}{\bibfnamefont{N.~J.} \bibnamefont{Cerf}},
  \bibinfo{author}{\bibfnamefont{T.~C.} \bibnamefont{Ralph}},
  \bibinfo{author}{\bibfnamefont{J.~H.} \bibnamefont{Shapiro}},
  \bibnamefont{and} \bibinfo{author}{\bibfnamefont{S.}~\bibnamefont{Lloyd}},
  \bibinfo{journal}{Reviews of Modern Physics} \textbf{\bibinfo{volume}{84}},
  \bibinfo{pages}{621} (\bibinfo{year}{2012}), ISSN \bibinfo{issn}{0034-6861,
  1539-0756}.

\bibitem[{\citenamefont{Lance et~al.}(2005)\citenamefont{Lance, Symul, Bowen,
  Sanders, Tyc, Ralph, and Lam}}]{lance_2005}
\bibinfo{author}{\bibfnamefont{A.~M.} \bibnamefont{Lance}},
  \bibinfo{author}{\bibfnamefont{T.}~\bibnamefont{Symul}},
  \bibinfo{author}{\bibfnamefont{W.~P.} \bibnamefont{Bowen}},
  \bibinfo{author}{\bibfnamefont{B.~C.} \bibnamefont{Sanders}},
  \bibinfo{author}{\bibfnamefont{T.}~\bibnamefont{Tyc}},
  \bibinfo{author}{\bibfnamefont{T.~C.} \bibnamefont{Ralph}}, \bibnamefont{and}
  \bibinfo{author}{\bibfnamefont{P.~K.} \bibnamefont{Lam}},
  \bibinfo{journal}{Physical Review A} \textbf{\bibinfo{volume}{71}},
  \bibinfo{pages}{033814} (\bibinfo{year}{2005}), ISSN
  \bibinfo{issn}{1050-2947, 1094-1622}.

\bibitem[{\citenamefont{Grosshans and Grangier}(2001)}]{grosshans_2001}
\bibinfo{author}{\bibfnamefont{F.}~\bibnamefont{Grosshans}} \bibnamefont{and}
  \bibinfo{author}{\bibfnamefont{P.}~\bibnamefont{Grangier}},
  \bibinfo{journal}{Physical Review A} \textbf{\bibinfo{volume}{64}},
  \bibinfo{pages}{010301(R)} (\bibinfo{year}{2001}).

\bibitem[{\citenamefont{Olivares et~al.}(2006)\citenamefont{Olivares, Paris,
  and Andersen}}]{olivares_2006}
\bibinfo{author}{\bibfnamefont{S.}~\bibnamefont{Olivares}},
  \bibinfo{author}{\bibfnamefont{M.~G.~A.} \bibnamefont{Paris}},
  \bibnamefont{and} \bibinfo{author}{\bibfnamefont{U.~L.}
  \bibnamefont{Andersen}}, \bibinfo{journal}{Physical Review A}
  \textbf{\bibinfo{volume}{73}}, \bibinfo{pages}{062330}
  (\bibinfo{year}{2006}), ISSN \bibinfo{issn}{1050-2947, 1094-1622}.

\bibitem[{\citenamefont{Gu{\c t}{\u a} and Matsumoto}(2006)}]{guta_2006}
\bibinfo{author}{\bibfnamefont{M.}~\bibnamefont{Gu{\c t}{\u a}}}
  \bibnamefont{and}
  \bibinfo{author}{\bibfnamefont{K.}~\bibnamefont{Matsumoto}},
  \bibinfo{journal}{Physical Review A} \textbf{\bibinfo{volume}{74}},
  \bibinfo{pages}{032305} (\bibinfo{year}{2006}), ISSN
  \bibinfo{issn}{1050-2947, 1094-1622}.

\end{thebibliography}

\end{document}


\title{Supplementary material: Quantum steering is the resource for secure quantum state sharing}
\author{Cailean Wilkinson, Matthew Thornton and Natalia Korolkova}
\affiliation{School of Physics and Astronomy, University of St Andrews,
North Haugh, St Andrews KY16 9SS, UK}
\maketitle

This Supplementary Material discusses further several of the results which are used in the main paper and presents their derivations. In \cref{supmat:dealer} we briefly motivate our choice of dealer protocol. In \cref{supmat:channel} we demonstrate the form the Gaussian channel for reconstructing the state using player 3's share takes. In \cref{supmat:example}, we then demonstrate that this channel can be implemented by a practical setup involving a feed-forward loop, and show some of the setup parameters for the reconstruction step. The corresponding result from \cref{supmat:channel} forms the basis of our quantitative results in the main paper. In \cref{supmat:modeswap} we demonstrate an interesting asymmetry present in this protocol where it is always preferable to swap the resource modes so the steering can be used for $g<1$, regardless of whether a better steering parameter is achievable in the other direction for $g>1$. Finally, we derive in full the results for general one-mode Gaussian secret states, which are described in section VI of the main paper. These are derived in \cref{supmat:squeezedstates} for general Gaussian pure states, and \cref{supmat:thermalstates} for general Gaussian mixed states.

\tableofcontents

\section{Choice of dealer protocol} \label{supmat:dealer}
A Quantum State Sharing protocol consists of two components: the initial dealer protocol in which the secret state is split between multiple shares, and the later reconstruction protocol in which it is recomposed from a subset of those shares. We begin here by briefly motivating our choice of QSS dealer setup. We will later in \cref{supmat:channel} show that the reconstruction protocol used in this Paper is optimal for this dealer setup.

There are a number of ways in which one could split the information describing a single quantum state between multiple modes. In the protocol we explore in this Paper, the dealer simply mixes the secret state with one mode of the resource state using a balanced beamsplitter. We will show here that this protocol is the optimal one, subject to the constraint that the dealer is limited to quantum channels defined by a real matrix,
\begin{align}
    \label{eqn:real_channels1}
    \mode_1 &= \alpha_1 \mode_\psi + \beta_1 \mode_\text{r1} + \gamma_1 \mode_\text{r2} \\
    \label{eqn:real_channels2}
    \mode_2 &=\alpha_2 \mode_\psi + \beta_2 \mode_\text{r1} + \gamma_2 \mode_\text{r2} \\
    \label{eqn:real_channels3}
    \mode_3 &=\alpha_3 \mode_\psi + \beta_3 \mode_\text{r1} + \gamma_3 \mode_\text{r2},
\end{align}
where $\alpha_i,\beta_i,\gamma_i \in \reals$ are such that the channel respects commutation relations, $\mode_{1,2,3}$ are the output (dealt) modes and $\mode_\psi, \mode_{\text{r}i}$ represent the secret and resource state modes respectively. This restricts the dealer to processes which mix $\x$ and $\p$ quadratures equally. In practice, this means restricting them to a protocol built up from beamsplitters, with no squeezers or nonlinear optics.

\subsection{Constraints placed on the dealt shares}
To find the set of dealer protocols which work for QSS, we first consider the conditions which define an optimal reconstruction. With that in mind, we impose the following three restrictions on the general dealer channel defined above in \cref{eqn:real_channels1,eqn:real_channels2,eqn:real_channels3}:
\begin{itemize}
    \item The dealer channel must, like all quantum channels, preserve the canonical commutation relations (CCRs).
    \item The original secret state must be reconstructable from any two outputs of the dealer protocol, and optimally we wish to minimise the information obtainable from any single output mode.
    \item It must always be possible to recombine the output modes such that the resource state cancels maximally, regardless of the specific resource used. This means the collaborating parties must be able to set up their reconstruction protocol to mix the resource mode contributions with ratio $g$ for any $g /in (0, /sqrt{2})$. 
    \item The impact of the protocol must be equivalent in each quadrature. This means that the output state must be able to be squeezed such that there is equal amplification on each quadrature, and that after such a squeezing there should not be more noise added to one quadrature than the other.
\end{itemize}

To consider the impact of these reconstruction conditions on the dealer protocol, we consider the effect of the reconstruction protocol on two arbitrary outputs of the dealer protocol. In general, the two shares used to reconstruct the state will then be of the form
\begin{align}
    \mode_1 &= \alpha_1 \mode_\psi + \beta_1 \mode_\text{r1} + \gamma_1 \mode_\text{r2} \\
    \mode_2 &=\alpha_2 \mode_\psi + \beta_2 \mode_\text{r1} + \gamma_2 \mode_\text{r2}.
\end{align}

The most general reconstruction channel can be written
\begin{align}
    \begin{pmatrix}
        \q_1 \\ \q_2  
    \end{pmatrix}
    \rightarrow
    \mat{T}
    \begin{pmatrix}
        \q_1 \\ \q_2
    \end{pmatrix}
\end{align}
for
\begin{align}
    \mat{T} = \begin{pmatrix}
        A^+ & 0 & B^+ & 0 \\
        0 & A^- & 0 & B^- \\
        C^+ & 0 & D^+ \\
        0 & C^- & 0 & D^- \\
    \end{pmatrix},
\end{align}
where $A^\pm,B^\pm,C^\pm,D^\pm \in \reals$.

The output from the reconstruction process is then
\begin{align}
    \q_{out} = (A^\pm \alpha_1 + B^\pm \alpha_2) \q_\psi + (A^\pm \beta_1 + B^\pm \beta_2) \q_{r1} +(A^\pm \gamma_1 + B^\pm \gamma_2) \q_{r2},
\end{align}
where we no longer consider the second mode as its form is irrelevant to our analysis.

Recalling that any reconstruction requires that the resource modes mix with some ratio $g$, we first reparameterise this output such that the coefficients for $\q_{r1}$ and $\q_{r2}$ differ by a factor of $\mp g$. This gives an expression for $B^\pm$ as
\begin{align}
    B^\pm = -\frac{\gamma_1 \pm g \beta_1 }{\gamma_2 \pm g \beta_2} A^\pm.
\end{align}

Imposing that the reconstruction preserve the CCRs (by the same method used in \cref{supmat:channel}) then provides expressions for $A^-$ (and $C^-$, $D^\pm$), leaving the channel fully parameterised by $A^+$ and $C^+$. The output state can then be written as
\begin{align}
    \q_{out} = \eta^\pm \left( \q_\psi + \lambda^\pm (\q_{r1} \mp g \q_{r2}) \right),
\end{align}
where
\begin{align}
    \eta^+ &= \frac{A^+ \left( \alpha_1 \gamma_2 - \alpha_2 \gamma_1 + g \left( \alpha_1 \beta_2 - \alpha_2 \beta_1 \right)\right)}{\gamma_2 +g \beta_2}, \\
    \eta^- &= -\frac{(\gamma_2 + g \beta_2) \left(\alpha_2 \gamma_1 - \alpha_1 \gamma_2 + g \left(\alpha_1\beta_2 - \alpha_2 \beta_1 \right)\right)}{A^+ \left(\gamma_1^2 + \gamma_2^2 - g^2 \left(\beta_1^2 + \beta_2^2 \right)\right)}, \\
    \lambda^\pm &= \frac{\beta_2 \gamma_1 - \beta_1\gamma_2}{\pm g (\alpha_2 \beta_1 - \alpha_1 \beta_2) + \alpha_2\gamma_1 - \alpha_1\gamma_2}.
\end{align}

The final channel parameter present, $A^+$, represents only a squeezing of the output state. In our analysis of the QSS protocol in the remainder of this Paper, we will further impose that the state is squeezed to equalise $\eta^+ = \eta^- := \eta(\alpha_i,\beta_i,\gamma_i)$, but for the remainder of this section the precise form of $\eta$ is not important.

For a given pair of input modes, this is the only channel which can reconstruct the original state $\mode_\psi$. We now ask which input modes allow for a reconstruction which meets the conditions we have set for our QSS protocol. Recall, we stipulated that the reconstruction must be quadrature-symmetric, in that it must not introduce more noise into one quadrature than the other once the quadratures have been squeezed to equal amplification. This is equivalent to stating that $\lambda^+ = \lambda^-$.

There are two ways this condition can be satisfied; the first is that
\begin{align}
    \label{eqn:dealercond1}
    \beta_2 \gamma_1 - \beta_1\gamma_2 = 0,
\end{align} 
in which case $\lambda^\pm = 0$ and there is no contribution from the resource state regardless of its steering. This is the case for the reconstruction protocol we use for shares 1 and 2, which uses a second beam splitter to exactly reversee the dealer protocol. 

The second option is that
\begin{align}
    \label{eqn:dealercond2}
    \alpha_2 \beta_1 - \alpha_1 \beta_2 = 0,
\end{align}
in which case
\begin{align}
    \lambda^\pm &= \frac{\beta_2 \gamma_1 - \beta_1\gamma_2}{\alpha_2\gamma_1 - \alpha_1\gamma_2}.
\end{align}

One of these two conditions must be satisfied for the original state to be suitably reconstructed from the two shares given.

\subsection{Dealer protocols satisfying these constraints}
We saw in the previous subsection that for the original state to be reconstructable from two shares in an optimal quadrature-symmetric way, one of the conditions in \cref{eqn:dealercond1,eqn:dealercond2} must be satisfied. In a quantum state sharing scheme, though, reconstruction must be possible for any permutation of 2 shares and so every permutation $i,j\in\{1,2,3\}$ of two outputs from the dealer protocol must satisfy
\begin{align}
    \beta_i \gamma_j = \beta_j \gamma_i, \quad \textit{or} \quad
    \alpha_i \beta_j = \alpha_j \beta_i.
\end{align}
These 3 conditions on the output of the dealer protocol, taken together with the 6 conditions placed on the dealer channel itself to preserve the CCRs, gives a total of 8 independent conditions on the parameters $\alpha_i$, $\beta_i$, $\gamma_i$ (with one superfluous condition).

There are a total of 3 possible dealer protocols satisfying these conditions that are distinct under mode relabelling and phase rotations. The output modes from these 3 protocols are given by
\begin{align}
    \begin{pmatrix}
        \mode_1 \\ \mode_2 \\ \mode_3
    \end{pmatrix}
    &= 
    \begin{pmatrix}
        1 & 0 & 0 \\
        0 & 1 & 0 \\
        0 & 0 & 1
    \end{pmatrix}
    \begin{pmatrix}
        \mode_\psi \\ \mode_\text{r1} \\ \mode_\text{r2}
    \end{pmatrix},
    \\
    \begin{pmatrix}
        \mode_1 \\ \mode_2 \\ \mode_3
    \end{pmatrix}
    &= 
    \begin{pmatrix}
        1 & 0 & 0 \\
        0 & \lambda & \sqrt{1-\lambda^2}  \\
        0 & \sqrt{1-\lambda^2} & - \lambda 
    \end{pmatrix}
    \begin{pmatrix}
        \mode_\psi \\ \mode_\text{r1} \\ \mode_\text{r2}
    \end{pmatrix},
    \\
    \begin{pmatrix}
        \mode_1 \\ \mode_2 \\ \mode_3
    \end{pmatrix}
    &=
    \begin{pmatrix}
    \lambda & \sqrt{1-\lambda^2} & 0 \\
    \sqrt{1-\lambda^2} & - \lambda & 0  \\
    0 & 0 & 1 
    \end{pmatrix}
    \begin{pmatrix}
        \mode_\psi \\ \mode_\text{r1} \\ \mode_\text{r2}
    \end{pmatrix}.
\end{align}

The first two of these options, in which the secret state is wholly contained in one output mode, are plainly unsuitable for QSS - there would be no obscuring of the secret mode. Consequently, we have only one class of beam splitter-based dealer protocol that is suitable for state sharing, parameterised by $\lambda\in[0,1]$.

The ideal QSS scheme is the one that minimises the amount of information obtainable from any single share, which is achieved here by setting $\lambda=1/\sqrt{2}$; setting $\lambda$ to any other value would decrease the contribution of the secret state to one share at the expense of increasing its contribution to the other. This would result in not only a worse reconstruction involving the share with a lower $\psi$ contribution, but also leave more information about $\psi$ vulnerable to observation from the other share. This is exactly the dealer protocol we have analysed in this Paper.

\subsection{Discussion}
We have shown here that the QSS protocol we have considered in this Paper is the optimal one when the dealer is limited to the use of beam splitters. In a real-world setting where quantum resources such as quadrature squeezers are relatively expensive, a wholly beam splitter dealer setup is likely to be preferable. In particular, we have also shown here that the asymmetry between the secret state contributions to modes 1 \& 2 and mode 3 is unavoidable. There is no beam splitter based dealer output which allows for the secret state to be split between all three modes which also allows it to be reconstructed from any two of them.

Were the dealer to have access to the wider selection of quantum operations, this could be modelled by replacing the real parameters in \cref{eqn:real_channels1,eqn:real_channels2,eqn:real_channels3} with complex parameters. Similar constraints could be derived for this case, though we would expect there to be more degrees of freedom left available for the dealer protocol and so a number of possible classes of dealer protocol to be available. Although this may add flexibility to the protocol, for example possibly allowing for the moving of quantum resources between the reconstruction and the dealer protocol, we do not expect this to be associated with an increase in reconstruction fidelity.

One question such an analysis would open up is whether there exists a dealer setup that completely removes the $g$ parameterisation from the state reconstruction step. This would allow for the collaborating players to reconstruct the original secret state with a setup completely blind to the specifics of the resource state used. The dealer would then have to adjust their protocol based on the specific resource state used. Such a protocol would allow for a single reconstruction setup to be used between multiple dealers who each use a different type of resource state.

\section{Quantum channel representing reconstruction involving player 3}\label{supmat:channel}
In this section, we derive the quantum channel representation of the state reconstruction protocol involving player $3$. The results from this section form the basis of all subsequent analyses of the protocol.

As described in section III of the main body of the paper, the optimum reconstruction process for shares 1 and 3 is represented by the operation
\begin{align}\label{eqn:reconstruction_13}
    \q_{\text{out}} &\rightarrow \eta \left(\sqrt{2}\q_1 \mp g\q_3 \right) \notag \\
    &= \eta \left( \q_\psi + (\q_{r1} \mp g\q_{r2}) \right),
\end{align}
where $g>0$ represents an adjustable experimental parameter and $\eta$ an as-yet unspecified gain on the output state. A similar operation exists for shares 2 and 3. We will show here that this gain must be of the form $\eta=1/\sqrt{2-g^2}$. While this operation can be represented by a Gaussian unitary, we will model it instead as a general Gaussian quantum channel to demonstrate that no more-efficient channel exists.

An arbitrary Gaussian quantum channel can be represented through its effect on the mean and covariance matrices of a Gaussian state as \cite{weedbrook_2012}
\begin{align}
    \mean \rightarrow \mat{T}\mean, \qquad \cov \rightarrow \mat{T}\cov\mat{T}^\mathrm{T}+\mat{N},
\end{align}
where $\mat{T}\in\reals^{2n}$ represents the mixing of the modes imparted by the channel and $\mat{N}\in\reals^{2n}$ represents additional environment noise introduced by the channel. For this channel to be trace-preserving and satisfy the uncertainty principle, it is required that
\begin{align}
    \mat{N} + i\mat{\Omega} - i\mat{T}\mat{\Omega}\mat{T}^\mathrm{T} \ge 0,
    \label{eqn:channel_condition}
\end{align}
where $\mat{O}\ge0$ represents the positive semi-definiteness condition that $\vec{z}^\dagger\mat{O}\vec{z} \ge 0 \,\, \forall\, \vec{z}\in\complex^{2n}$ and $\mat{\Omega}$ is the standard symplectic form representing the canonical commutation relations
\begin{align}
    \mat{\Omega} = \bigoplus_{n}
    \begin{pmatrix}
        0 & 1 \\
        -1 & 0
    \end{pmatrix}.
\end{align}

The operation described by \cref{eqn:reconstruction_13} can be represented as
\begin{align}
    \mat{T} = 
    \begin{pmatrix}
        \sqrt{2}\eta & 0 & -g\eta & 0 \\
        0 & \sqrt{2}\eta & 0 & g\eta \\
        a & 0 & c & 0 \\
        0 & b & 0 & d
    \end{pmatrix},
\end{align}
where $a,b,c,d\in\reals$ represent the second output of this channel and are left unspecified and we have set $\mat{N}=0$ to remove unnecessary environment noise sources. The condition for this channel to be physical then becomes
\begin{align}
    i\mat{\Omega} - i\mat{T}\mat{\Omega}\mat{T}^\mathrm{T} = i 
    \begin{pmatrix}
        0 & g^2 \eta ^2-2 \eta ^2+1 & 0 & \eta (d g -\sqrt{2} b) \\
        -g^2 \eta ^2+2 \eta ^2-1 & 0 & \eta(\sqrt{2} a +c g) & 0 \\
        0 & \eta(-\sqrt{2} a -c g) & 0 & -a b-c d+1 \\
        \eta( \sqrt{2} b -d g) & 0 & a b+c d-1 & 0 \\
    \end{pmatrix} \ge 0.
\end{align}

Taking an arbitrary $\vec{z}\in\complex^4$, which we write as $\vec{x}+i\vec{y}$ for $\vec{x},\vec{y}\in\reals^4$, we can calculate
\begin{align}
    \vec{z}^\dagger(i\mat{\Omega} - i\mat{T}\mat{\Omega}\mat{T}^\mathrm{T})\vec{z}
    \nonumber
    =\, &x_1 \left(
                2 \left((2-g^2)\eta^2 - 1 \right) y_2
                + 2\eta \left( \sqrt{2}b - dg \right) y_4
            \right) \\
    \nonumber
    -\, &x_2 \left(
                2 \left((2 - g^2)\eta^2 - 1 \right) y_1
                + 2\eta \left(\sqrt{2}a + cg \right) y_3 
            \right) \\
    \nonumber
    +\, &x_3  \left(
                2\eta \left( \sqrt{2}a + cg \right) y_2
                + 2 \left( ab + cd - 1 \right) y_4
            \right) \\
    -\, &x_4 \left(
                2\eta \left( \sqrt{2}b - dg \right) y_1
                + 2\left( ab + cd - 1 \right) y_3
            \right).
\end{align}
Clearly, for this expression to be non-negative for all $\mat{z}\in\complex^4$ the coefficients of each permutation of $x_iy_j$ must be individually zero. Hence, for $\mat{T}$ to represent a gaussian channel, the following four conditions must be satisfied.
\begin{align}
    \sqrt{2}b - d g &= 0, \\
    \sqrt{2}a + c g &= 0, \\
    1 - ab - cd &= 0, \\
    1 - \eta^2 (2 - g^2) &= 0.
\end{align}
The first three of these conditions define the form of the second output of the channel as 
\begin{align}
    b &= \frac{g^2}{a (g^2 - 2)}, \\
    c &= - \frac{\sqrt{2}}{g} a, \\
    d &= \frac{\sqrt{2}g}{a(g^2-2)},
\end{align}
for any $a\in\reals$. The final condition gives the gain which the channel must impart on the secret state as
\begin{align}
    \eta = \frac{1}{\sqrt{2-g^2}}.
\end{align}
We show in the next section that this gain $\eta$ is reproduced by a physical setup implementing the operation in \cref{eqn:reconstruction_13}.

\section{An example \{1,3\} reconstruction setup}\label{supmat:example}
\begin{figure}
    \includegraphics[width=\linewidth]{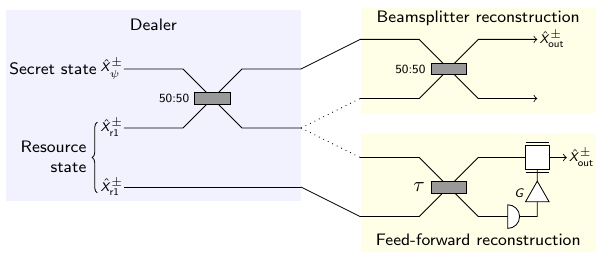}
    \caption{One possible $(2,3)$-threshold QSS setup, sketching $\{1,2\}$-reconstruction and $\{2,3\}$-reconstruction. The $\{1,3\}$-reconstruction scheme is identical to $\{2,3\}$-reconstruction. The three shares are produced by the dealer by mixing the secret state with one mode of a two-mode resource state. Shares 1 and 2 can be combined in a second balanced beamsplitter to reconstruct the original state. Shares 1 and 3 or 2 and 3 can be used to reconstruct the state with a feed-forward process. The shares are first mixed on a beamsplitter with transmissivity $\tau$ to get the desired mixing in one quadrature, and then a feed-forward process is used to obtain the desired mixing in the other quadrature.}
    \label{fig:ff_setup}
\end{figure}
In this section we discuss one possible implementation of the $\{1,3\}$ reconstruction operation required for \cref{eqn:reconstruction_13}. A number of different implementations have been proposed for this operation, and a full discussion of each of these for the unity-gain $g=1$ case can be found in Ref. \cite{lance_2005}. Any of these, with suitable adjustments, can be used for arbitrary $g\in(0,\sqrt{2})$ QSS. 

\subsection{Dealer protocol}
The dealer process, \cref{fig:ff_setup} is identical regardless of the physical setup used to reconstruct the secret. The secret state is passed through a balanced beamsplitter with one mode of the 2-mode resource state to produce three shares of the form
\begin{align}
    \q_1 &= \frac{1}{\sqrt{2}} ( \q_\psi + \q_{r1} ), \\
    \q_2 &= \frac{1}{\sqrt{2}} ( \q_\psi - \q_{r1} ), \\
    \q_3 &= \q_{r2},
\end{align}
where $\q_\psi$ represents the secret state and $\q_{ri}$ the two modes of the resource.

\subsection{State reconstruction through single feed-forward process}
The difficulty in implementing the operation from \cref{eqn:reconstruction_13} lies in the opposing mixing in the $\x$ and $\p$ quadratures - which is to say that the output quadratures have components $\x_1 - g\x_2$ and $\p_1 + g\p_2$. The approach taken in this implementation, shown in \cref{fig:ff_setup}, is to first use an unbalanced beamsplitter to get the correct mixing in one quadrature and then use a single-quadrature digital feed-forward step to correct the other quadrature.

After the beamsplitter the two shares are combined such that
\begin{align}
    \q_A &= \sqrt{\frac{\tau}{2}} \left( \q_\psi + (\q_{r1} - \sqrt{\frac{2-2\tau}{\tau}} \q_{r2} ) \right), \\
    \q_B &= \sqrt{\frac{1-\tau}{2}} \left( \q_\psi + (\q_{r1} + \sqrt{\frac{2\tau}{1-\tau}} \q_{r2} ) \right),
\end{align}
where $\tau\in[0,1]$ represents the transmissivity of the beamsplitter. By selecting $\tau$ such that $\sqrt{(2-2\tau)/\tau}=g$, which is always possible for the range $g\in(0,\sqrt{2})$ for which steering is possible, one can see that $\q_A$ is proportional to the desired form of \cref{eqn:reconstruction_13} in the $\x$ quadrature.

The $\p_B$ quadrature is then measured and the $\p_A$ quadrature displaced by the corresponding value with a gain $G\in\reals$, giving
\begin{align}
    \p_A &\rightarrow \p_A + G \p_B \\
    &= \sqrt{\frac{\tau}{2}} \left( \p_\psi + (\p_{r1} - \sqrt{\frac{2-2\tau}{\tau}} \p_{r2} ) \right) + G \sqrt{\frac{1-\tau}{2}} \left( \p_\psi + (\p_{r1} + \sqrt{\frac{2\tau}{1-\tau}} \p_{r2} ) \right) \\
    &= \big( \sqrt{\frac{\tau}{2}}+ G \sqrt{\frac{1-\tau}{2}}\big) \bigg( \p_\psi + \p_{r1} + \sqrt{2}\frac{G\sqrt{\tau} - \sqrt{1-\tau}}{\sqrt{\tau}+G\sqrt{1-\tau}} \p_{r2} \bigg),
\end{align}
with $\x_A$ left unaltered. Selecting $G$ such that
\begin{align}
    \sqrt{2}\frac{G\sqrt{\tau} - \sqrt{1-\tau}}{\sqrt{\tau}+G\sqrt{1-\tau}} = g,
\end{align}
which is again always possible for any $g\in(0,\sqrt{2})$, this output state becomes
\begin{align}
    \x_\text{out} &=\frac{1}{\sqrt{2+g^2}} ( \x_\psi + \x_{r1} - g \x_{r2} ),
    \\ 
    \p_\text{out} &= \frac{\sqrt{2+g^2}}{2-g^2} ( \p_\psi + \p_{r1} + g \p_{r2} ). 
\end{align}

\noindent By applying an ideal squeezing operation to this state, we can equalise the pre-factors and reach an output in the same form as expected from the previous section,
\begin{align}
    \q_\text{out} = \frac{1}{\sqrt{2-g^2}} \left( \q_\psi + (\q_{r1} \mp g\q_{r2}) \right).
\end{align}

\noindent The values of $\tau$ and $G$ which correspond to a given $g\in(0,\sqrt{2})$ are
\begin{align}
    \tau &= \frac{2}{2+g^2} \\
    G &= \frac{2\sqrt{2}g}{2-g^2}.
\end{align}

\section{Swapping resource modes}\label{supmat:modeswap}
In this section we demonstrate that while QSS can be implemented for any $g\in(0,\sqrt{2})$, it is always preferable to use $g\le1$ even if a resource state would be better steerable for $g>1$. This can be achieved in most setups through the dealer swapping the labelling of the modes. The results of this section underpin Result~2 of the main paper.

The key point in our argument is that any state steerable in one direction for $g>1$ with steering parameter $\steering$ is also necessarily steerable in the other direction for $\bar{g}:=1/g$ with steering parameter
\begin{align}
    \label{eqn:mode_swapping}
    E_{2|1}(\bar{g}) = \frac{1}{g^2} E_{1|2}(g).
\end{align}
This can be shown trivially by expanding the steering parameter $\steering = \Delta^2(\q_1 \mp g \q_2)$ in terms of the variance of each mode and the covariance between them in each case.

Let us consider then QSS implemented for $g>1$ for a given resource with steering $\steering$. Using the general output derived later in this document in \cref{sec:thermal}, the state will then be reconstructed with fidelity
\begin{align}
    \mathcal{F}_{1|2} = 2/\sqrt{4 + 4(2\bar{n}+1)\cosh(2\zeta)(1+\steering-1/\eta^2) + (2\bar{n}+1)^2(1+\steering-1/\eta^2)^2},
\end{align}
for $\eta=1/\sqrt{2-g^2}$.

Should the dealer decide instead to swap the resource modes and utilise the steering in the opposite direction for $\bar{g}$, then the state would be reconstructed with fidelity
\begin{align}
    \mathcal{F}_{2|1} = 2/\sqrt{4 + 4(2\bar{n}+1)\cosh(2\zeta)(1+\bar{\eta}^2E_{2|1}(\bar{g})-\bar{\eta}^2) + (2\bar{n}+1)^2(1+\bar{\eta}^2E_{2|1}(\bar{g})-\bar{\eta}^2)^2},
\end{align}
for $\bar{\eta}=1/\sqrt{2-\bar{g}^2}$.

Noting that $\bar{n}\ge0$ and $\cosh(2\zeta)\ge1$, it is clear that $\mathcal{F}_{2|1} > \mathcal{F}_{1|2}$ if and only if
\begin{align}
    1+\bar{\eta}^2E_{2|1}(\bar{g})-\bar{\eta}^2 < 1+\steering-1/\eta^2.
\end{align}

Recalling \cref{eqn:mode_swapping} and that $\bar{g} = 1/g$, we can rewrite this condition as
\begin{align}
    \frac{1+\steering-1/\eta^2}{2g^2-1} < 1+\steering-1/\eta^2,
\end{align}
which is trivially satisfied for all $g>1$.

Consequently, whenever a resource state is optimally steerable in one direction for $g>1$, it is always preferable to swap the order of the modes and instead utilise the steering in the opposite direction.

\section{QSS for general pure Gaussian states: squeezed coherent states} \label{supmat:squeezedstates}
We now turn our attention to the use of this quantum state sharing protocol for a wider class of Gaussian states. In this section we will consider the most general pure Gaussian state by introducing an unknown degree of squeezing to the coherent states we have considered previously. The results from this section form the basis of section VI.A in the main paper.

Our starting point is the squeezed coherent state with arbitrary mean $\mean\in\reals^2$ and covariance matrix
\begin{align}
    \cov = \begin{pmatrix}
        \e^{-2\zeta}\cos\theta + e^{2\zeta}\sin\theta & 2\sinh(2\zeta)\cos\theta\sin\theta \\
        2\sinh(2\zeta)\cos\theta\sin\theta & \e^{2\zeta}\cos\theta + e^{-2\zeta}\sin\theta
    \end{pmatrix},
\end{align}
where $\zeta\in\reals$ represents the degree to which the state is squeezed, and $\theta$ the angle in phase space along which it is squeezed. As the entire QSS protocol is phase independent, we will in this paper assume without loss of generality that the secret state is squeezed along $\theta=0$, resulting in a covariance matrix of
\begin{align}
    \label{eqn:squeezed_cov}
    \cov_\psi = \begin{pmatrix}
        \e^{-2\zeta} & 0 \\
        0 & \e^{2\zeta}
    \end{pmatrix}.
\end{align}

A the protocol has not been changed from that discussed previously for coherent states, the ideal output is still that described in section IV of the main body of the paper,
\begin{align}
    \q_\text{out} = 
    \begin{cases}
        \q_\psi + \q_{r1} \mp g\q_{r2} + \sqrt{\frac{1}{\eta^2}-1} \q_\text{vac} & g <= 1 \\        
        \q_\psi + \q_{r1} \mp g\q_{r2} + \sqrt{1-\frac{1}{\eta^2}} \q_\text{vac} & g >= 1
    \end{cases}
\end{align}
where $\q_\psi$ represents contribution from the secret state, $\q_{ri}$ that of the two modes of the resource state, $\q_\text{vac}$ an auxiliary vacuum mode, and $\eta = 1/\sqrt{2-g^2}$ as previously. The mean of this output state is then exactly the mean of the input state, $\mean_\text{out} = \mean_\psi$, while the covariance matrix is given by
\begin{align}
    \cov_\text{out} &= \cov_\psi +
    \begin{cases}
        (\eta^2 E_{1|2}(g) + 1 - \eta^2)\mat{I} & g \le 1 \\
        (E_{1|2}(g) + 1 - 1/\eta^2)\mat{I} & g \ge 1
    \end{cases},
    \label{eqn:pure_output_cov}
\end{align}
with $\cov_\psi$ the squeezed state covariance matrix given in \cref{eqn:squeezed_cov}.

\subsection{Impact of QSS on squeezing parameter}
Before we discuss the usual metrics for quantifying the efficacy of this protocol for squeezed states, we first briefly consider the impact it has on the degree to which the secret state is squeezed before and after the protocol's application.

Consider first the case of \textit{esa} QSS, when $g\le1$. In this case, from \cref{eqn:pure_output_cov}, we know the reconstructed secret state has covariance matrix given by
\begin{align}
    \cov_\text{out} &= 
    \begin{pmatrix}
        \e^{-2\zeta} & 0 \\
        0 & \e^{2\zeta} 
    \end{pmatrix}
    +
    \begin{pmatrix}
        \eta^2 E_{1|2}(g) + 1 - \eta^2 & 0 \\
        0 & \eta^2 E_{1|2}(g) + 1 - \eta^2
    \end{pmatrix}
\end{align}
the original squeezed state mixed with an unsqueezed thermal state. This is equivalent to a thermal state which has been squeezed along the same angle to some degree $\zeta^\prime\in\reals$,
\begin{align}
    (2\tilde{n}+1)
    \begin{pmatrix}
        \e^{-2\zeta^\prime} & 0 \\
        0 & \e^{-2\zeta^\prime}
    \end{pmatrix}.
\end{align}
Equating these two representations we find that the output state is a squeezed thermal state with squeezing parameter
\begin{align}
    \zeta^\prime = \frac{1}{4} \ln\left[ \frac{\e^{2\zeta} + \eta^2\steering + 1 - \eta^2}{\e^{-2\zeta} + \eta^2\steering + 1 - \eta^2} \right] < \zeta,
\end{align}
which is strictly less than the squeezing of the input state $\zeta$.

Similarly, for \textit{lsatt} QSS, when $g\ge1$, we find that the output state is a squeezed thermal state with squeezing parameter $\zeta^\prime$ given by
\begin{align}
    \zeta^\prime &= \frac{1}{4} \ln\left[ \frac{\e^{2\zeta} + \steering + 1 - \frac{1}{\eta^2}}{\e^{-2\zeta} + \steering + 1 - \frac{1}{\eta^2}} \right] < \zeta,
\end{align}
in which the degree of squeezing has again been reduced below that of the input state.

\subsection{Fidelity and security conditions for squeezed state QSS}
\begin{figure}
    \subfloat[][]{
        \includegraphics[width=0.3\textwidth]{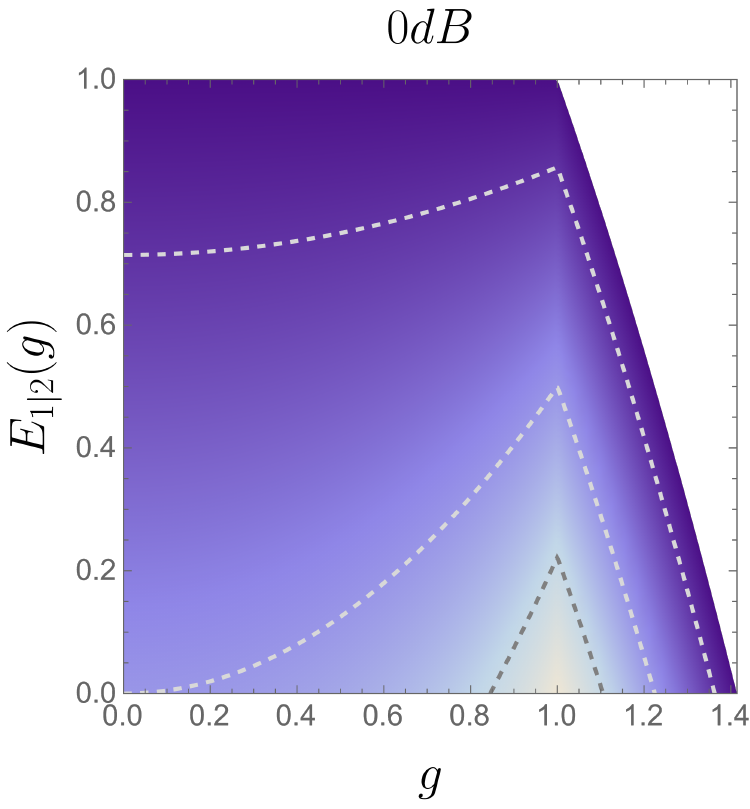}
    }
    \subfloat[][]{
        \includegraphics[width=0.3\textwidth]{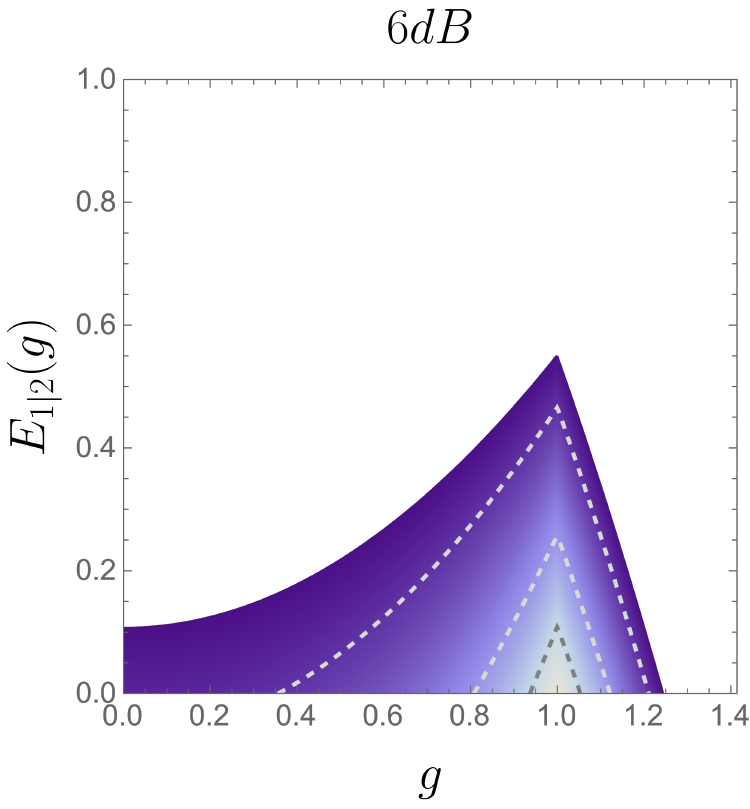}
    }
    \subfloat[][]{
        \includegraphics[width=0.35\textwidth]{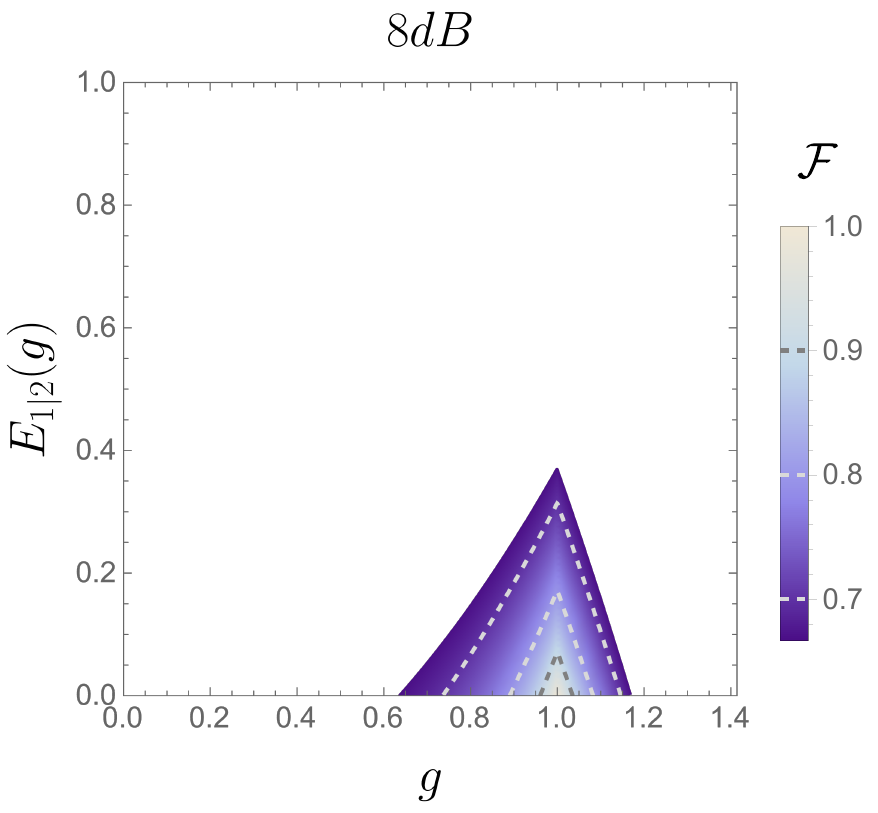}
    }
    \caption{Reconstruction fidelity for increasing resource steering for pure Gaussian states with (a) $0dB$ (no squeezing), (b) $6dB$, (c) $8dB$ squeezing levels respectively. The colour denotes the best attainable reconstruction fidelity when QSS is implemented for $g$ using a resource state with steering parameter $\steering$. The dashed lines denotes $\mathcal{F}=0.7,0.8,0.9$, and only the secure region $\mathcal{F}>2/3$ is shown. The region of allowable resource states shrinks as the squeezing of the secret state increases.
    }
    \label{fig:pure}
\end{figure}

Before we consider the achievable fidelity for the use of this protocol for squeezed states, we first discuss the question of security for such states. The accepted approach to security for quantum communication protocols is to consider the security guaranteed when it is impossible for a bad actor to gain more information about the transmitted state than the intended recipients. For coherent states, it has been shown that the optimal fidelity when cloning unknown states is $\mathcal{F}=2/3$ \cite{grosshans_2001}. Should one party reconstruct the state with a greater fidelity than this bound, it is then guaranteed that no other state containing more information than their reconstruction can exist. This is therefore the security condition we have considered thus far.

For states with unknown squeezing, though, this bound is not saturated; the optimal protocol for coherent state cloning is unable to reach $\mathcal{F}=2/3$ fidelity for squeezed states and the optimal cloning fidelity is not yet known \cite{olivares_2006}. Any Gaussian cloning method which allows for unknown squeezing still cannot, however, exceed the $\mathcal{F}=2/3$ threshold. In the absence of a tighter bound, we will continue to use this threshold as our security guarantee in this Paper. We note that reaching this bound is still a sufficient condition for security, but it may not always be necessary.

\subsubsection{\textit{esa} QSS for pure Gaussian states}
We first explore the case when $g\le1$ ($\eta\le1$) in which the secret state is pre-amplified before being attenuated by the QSS protocol. From \cref{eqn:pure_output_cov}, we can calculate the reconstruction fidelity simply as
\begin{align}
    \mathcal{F} &= \frac{2}{\sqrt{\det(\cov_\text{in} + \cov_\text{out})}} \\
    &= \frac{2}{\sqrt{(2\e^{2\zeta}+(1-\eta^2)+\eta^2\ent)(2\e^{-2\zeta}+(1-\eta^2)+\eta^2\ent)}}.
    \label{eqn:pure_esa_fidelity}
\end{align}
The effect of squeezing on the reconstruction fidelity for pure Gaussian states is shown in \cref{fig:pure}.

As discussed, we will consider the QSS protocol secure when the fidelity exceeds the usual coherent-state no-cloning threshold of $\mathcal{F}=2/3$. Applying this to \cref{eqn:pure_esa_fidelity} we can derive a condition on the amount of steering required to guarantee this protocols security when sharing an arbitrary squeezed state of squeezing up to $\zeta$ as
\begin{align}
    \mathcal{F} > \frac{2}{3}
    &\implies \frac{4}{(2\e^{2\zeta}+(1-\eta^2)+\eta^2\ent)(2\e^{-2\zeta}+(1-\eta^2)+\eta^2\ent)} > \frac{4}{9} \\
    &\implies (2\e^{2\zeta}+(1-\eta^2)+\eta^2\ent)(2\e^{-2\zeta}+(1-\eta^2)+\eta^2\ent) < 9 \\
    &\implies \eta^2\ents + 2\ent \left[ 2\cosh(2\zeta)+1-\eta^2 \right] + \left[\eta^2-\frac{4}{\eta^2}+2\cosh(2\zeta)(\frac{1}{\eta^2}-1)-2 \right] < 0.
\end{align}
Solving for $\ent$ then gives the security condition
\begin{align}
    \ent < 1 - \frac{1}{\eta^2} \left[ 1 + 2\cosh(2\zeta) - \sqrt{4\cosh^2(2\zeta)+5} \right]
\end{align}

\subsubsection{\textit{lsatt} QSS for pure Gaussian states}
For the case when $g\ge1$ ($\eta\ge1$) we can calculate the fidelity as
\begin{align}
    \mathcal{F} &= \frac{2}{\sqrt{\det(\cov_\text{in} + \cov_\text{out})}} \\
    &= \frac{2}{\sqrt{(2\e^{2\zeta}+(1-1/\eta^2)+\ent)(2\e^{-2\zeta}+(1-1/\eta^2)+\ent)}}.
\end{align}
This reconstruction fidelity for squeezed states is again shown in \cref{fig:pure}.

\noindent
Again, imposing the security condition of $\mathcal{F}>2/3$ gives the steering condition
\begin{align}
    (2\e^{2\zeta}+(1-1/\eta^2)+\ent)(2\e^{-2\zeta}+(1-1/\eta^2)+\ent) < 9, \\
    \implies \ents + 2\ent \left[ 2\cosh(2\zeta) + 1 - \frac{1}{\eta^2} \right] + \left[ \frac{1}{\eta^4} - \frac{2}{\eta^2} + 2\cosh(2\zeta)(1-\eta^2) - 4 \right] < 0,
\end{align}
and solving for $\ent$ gives
\begin{align}
    \ent < \frac{1}{\eta^2} - \left[ 1 + 2\cosh(2\zeta) - \sqrt{4\cosh(2\zeta)+5} \right].
\end{align}

\subsubsection{Overall security condition}
Taking these two cases together, we can then express the security condition for a squeezed state QSS protocol for any two-mode Gaussian resource state as 
\begin{result}
    A QSS protocol for the sharing of pure Gaussian secret state with squeezing of up to $\zeta_\text{max}$ is secure if the resource state used has steering of
    \begin{align}
        \steering < 
        \begin{cases}
            1 - \frac{1}{\eta^2} \Gamma(\zeta_\text{max}) & g \le 1 \quad(\eta\le1) \\
            \frac{1}{\eta^2} - \Gamma(\zeta_\text{max}) & g \ge 1 \quad(\eta\ge1)
        \end{cases},
    \end{align}
    for some $g\in(0,\sqrt{2})$ where $1/\eta^2 = 2-g^2$ and
    \begin{align}
        \Gamma(\zeta) = 1 + 2\cosh(2\zeta) - \sqrt{4\cosh^2(2\zeta)+5} \ge 0
    \end{align}
    is a monotonically increasing function of $\zeta$ with $\Gamma(0)=0$,
\end{result}

\noindent
which gives result 3 of the main paper.

\section{QSS for general mixed Gaussian states: squeezed thermal states} \label{sec:thermal} \label{supmat:thermalstates}

Finally, we consider the use of this protocol for any single-mode Gaussian state by allowing for thermal noise in the squeezed state we considered in the previous section. The results from this section form the basis for section VI.C of the main paper. The secret input state considered in this section is then the squeezed thermal state with arbitrary mean $\mean\in\reals^2$ and covariance matrix
\begin{align}
    \label{eqn:thermal_cov}
    \cov_\psi = (2 \bar{n} + 1) 
    \begin{pmatrix}
        \e^{-2\zeta} & 0 \\
        0 & \e^{2\zeta}
    \end{pmatrix},
\end{align}
where $\bar{n}\ge0$ represents the average thermal photon number in the state and $\zeta\in\reals$ represents the squeezing parameter and where we have again neglected the squeezing angle.

\subsection{Fidelity measure for Gaussian mixed states}
As we are now considering Gaussian states which are not, in general, pure the expression for fidelity we have used in previous sections no longer applies. Instead, we consider the full form of the fidelity measure for mixed states \cite{weedbrook_2012},
\begin{align}
    \mathcal{F} = \frac{2}{\sqrt{\Delta + \delta} - \sqrt{\delta}} 
    \e^{-\frac{1}{2} (\mean_\psi - \mean_\text{out})^\T (\cov_\psi + \cov_\text{out})^{-1} (\mean_\psi - \mean_\text{out})},
\end{align}
where
\begin{align}
    \Delta &= \det(\cov_\psi + \cov_\text{out}), \\
    \delta &= (\det\cov_\psi - 1) (\det\cov_\text{out} - 1).
\end{align}

\noindent
In this case, as the output state has been normalised so that $\mean_\text{out} = \mean_\psi$, we can neglect the exponential component the fidelity can be found simply as
\begin{align}
    \mathcal{F} = \frac{2}{\sqrt{\Delta + \delta} - \sqrt{\delta}}.
\end{align}

\subsection{Fidelity calculation for QSS state reconstruction}
As we have not altered the protocol from the coherent state case, the output state continues to have the form
\begin{align}
    \cov_\text{out} = \cov_\psi + \chi \mat{I},
\end{align}
where $\chi$ represents noise added from the use of an imperfect resource state,
\begin{align}
    \chi = \begin{cases}
        \eta^2 \steering + 1 - \eta^2 & g \le 1 \quad \textit{(esa)} \\
        \steering + 1 - \frac{1}{\eta^2} & g \ge 1 \quad \textit{(lsatt)}
    \end{cases},
\end{align}
and $\cov_\psi$ is the thermal state covariance matrix shown in \cref{eqn:thermal_cov}.

Thus,
\begin{align}
    \Delta &= \det(\cov_\psi + \cov_\text{out}) \\
    &= \det(2\cov_\psi + \chi \mat{I}) \\
    &= \left( 2 \tilde{n} \e^{-2\zeta} + \chi \right) \left( 2 \tilde{n} \e^{2\zeta} + \chi \right) \\
    &= 4 \tilde{n} + \chi^2 + 4\tilde{n} \chi \cosh({2\zeta}), \\
    \delta &= (\det\cov_\psi - 1) (\det\cov_\text{out} - 1) \\
    &= (\tilde{n}^2 - 1) \big( (\tilde{n} \e^{-2\zeta} + \chi ) ( \tilde{n} \e^{2\zeta} + \chi ) - 1 \big) \\
    &= (\tilde{n}^2 - 1) \left( \tilde{n}^2 + \chi^2 + 2 \tilde{n} \chi \cosh(2\zeta) - 1 \right),
\end{align}
and so
\begin{align}
    \Delta + \delta &= 4 \tilde{n} + \chi^2 + 4\tilde{n} \chi \cosh({2\zeta}) + (\tilde{n}^2 - 1) \left( \tilde{n}^2 + \chi^2 + 2 \tilde{n} \chi \cosh(2\zeta) - 1 \right) \\
    &= \chi^2 \tilde{n}^2 + 2 \tilde{n} \chi \cosh(2\zeta) + 2 \tilde{n}^2 \chi \cosh(2\zeta) + \tilde{n}^2 + 2 \tilde{n}^2 + 1 \\
    &= \left( \tilde{n} \chi + (\tilde{n}^2 + 1) \e^{2\zeta} \right) \left( \tilde{n} \chi + (\tilde{n}^2 + 1) \e^{-2\zeta} \right).
\end{align}

The fidelity for the thermal state reconstruction is then
\begin{align}
    \mathcal{F} &= \frac{2}{\sqrt{\Delta + \delta} - \sqrt{\delta}} \\
    &= \frac{2}{ \sqrt{\left( \tilde{n} \chi + (\tilde{n}^2 + 1) \e^{2\zeta} \right) \left( \tilde{n} \chi + (\tilde{n}^2 + 1) \e^{-2\zeta} \right)} - \sqrt{(\tilde{n}^2 - 1) \left( \tilde{n}^2 + \chi^2 + 2 \tilde{n} \chi \cosh(2\zeta) - 1 \right)}},
\end{align}
where $\tilde{n} = (2\bar{n}-1)$ represents the thermal state variance and 
\begin{align}
    \chi = \begin{cases}
        \eta^2 \steering + 1 - \eta^2 & g \le 1 \quad \textit{(esa)} \\
        \steering + 1 - \frac{1}{\eta^2} & g \ge 1 \quad \textit{(lsatt)}
    \end{cases}.
\end{align}

\noindent
In the case of pure Gaussian states, in which $\bar{n} = 0$ and so $\tilde{n} = 1$, the second root in the denominator disappears and the fidelity reduces to that previously derived in supplemental \cref{supmat:squeezedstates}. The achievable fidelity for thermal state QSS is shown in \cref{fig}.

\subsection{Security}
In contrast to previous sections, we do not present a security analysis here for thermal state QSS. As they do not saturate the uncertainty limit, thermal states may be cloned with significantly greater fidelity than their coherent state counterparts. In the limit of infinite thermal photon number they can be cloned perfectly. Although some work has been done into deriving fidelity bounds for the cloning of thermal states, this has not been shown to be optimal \cite{olivares_2006}, or has been shown to be optimal only for minimising the norm distance between the clones and the original state \cite{guta_2006}. Should an optimal expression for the cloning fidelity of thermal states be found, then a security condition could be derived for a subset of the thermal states by a similar method used here for coherent and squeezed states.

\bibliography{bibliography.bib}